\documentclass{jfm}

\usepackage{graphicx}
\usepackage{newtxtext}
\usepackage{newtxmath}
\usepackage{natbib}
\usepackage{hyperref}
\hypersetup{
    colorlinks = true,
    urlcolor   = blue,
    citecolor  = blue,
}

\newcommand{\RomanNumeralCaps}[1]
\linenumbers

\usepackage{siunitx}
\usepackage{amssymb}

\title{Turbidity currents propagating down an inclined slope: particle auto-suspension}

\author{Jiafeng Xie\aff{1},
  Peng Hu\aff{1,4}
  \corresp{\email{pengphu@zju.edu.cn}},
  Chenlin Zhu\aff{2}
  \corresp{\email{zhuclgary@foxmail.com}},
  Zhaosheng Yu\aff{3},
  and Thomas P\"ahtz\aff{1,5}}

\affiliation{\aff{1}Ocean College, Zhejiang University, Zhoushan 316021, PR China
\aff{2}Key Laboratory of Intelligent Manufacturing Quality Big Data Tracing and Analysis of Zhejiang Province, China Jiliang University, Hangzhou 310018, PR China
\aff{3}State Key Laboratory of Fluid Power and Mechatronic Systems, Department of Mechanics, Zhejiang University, Hangzhou 310027, PR China
\aff{4}Ocean Research Center of Zhoushan, Zhejiang University, Zhoushan 316021, PR China
\aff{5}Donghai Lab, Zhoushan 316021, PR China
}

\begin{document}
\maketitle

\begin{abstract}
The Turbidity current (TC), a ubiquitous fluid-particle coupled phenomenon in the natural environment and engineering, can transport over long distances on an inclined terrain due to the suspension mechanism. A large-eddy simulation and discrete element method coupled model is employed to simulate the particle-laden gravity currents over the inclined slope in order to investigate the auto-suspension mechanism from a Lagrangian perspective. The particle Reynolds number in our TC simulation is $0.01\sim0.1$ and the slope angle is $1/20 \sim 1/5$. The influences of initial particle concentration and terrain slope on the particle flow regimes, particle movement patterns, fluid-particle interactions, energy budget and auto-suspension index are explored. The results indicate that the auto-suspension particles predominantly appear near the current head and their number increases and then decreases during the current evolution, which is positively correlated with the coherent structures around the head. When the turbidity current propagates downstream, the average particle Reynolds number of the auto-suspension particles remains basically unchanged, and is higher than that of other transported particles. The average particle Reynolds number of the transported particles exhibits a negative correlation with the Reynolds number of the current. Furthermore, the increase in particle concentration will enhance the particle velocity, which allows the turbidity current to advance faster and improves the perpendicular support, thereby increasing the turbidity current auto-suspension capacity. Increasing slope angle will result in a slightly larger front velocity, while the effect of that on the total force is insignificant.
\end{abstract}

%\begin{keywords}
% Turbidity current, auto-suspension, head average forces, LES-DEM model.
%\end{keywords}

\section{\label{sec:Introducion}Introduction}

Turbidity currents (TCs) have been recognized as a widespread geophysical phenomenon in, for example, oceans, reservoirs, estuaries and lakes, which can transport large numbers of deposited particles long distances downstream \citep{middleton1966experiments,meiburg2010turbidity,wells2021turbulence}. With large-scale and high-intensity sediment transport, TC is inextricably linked with the rapid changes in riverbed morphology \citep{parker1987experiments}, the formation of submarine oil and gas \citep{meiburg2015modeling} and the stability of underwater structures \citep{fine2005grand}, and has attracted the attention of many researchers. 

In recent years, researchers \citep{simpson1982gravity,meiburg2015modeling,ouillon2021gravity} stated that it is the horizontal pressure difference between the muddy and ambient fluid area, caused by the particle suspension, which drives the horizontal movement of the flow. The suspension regime in TC can be divided into two types: re-suspension and auto-suspension. Both of them are essentially the upward particle suspension in the bed-normal direction. Re-suspension is equivalent to erosion and refers to particles being lifted from the deposited substrate by turbulence \citep{meiburg2010turbidity,strauss2012turbidity}, while auto-suspension corresponds to the event that the particles could remain suspended without additional fluid net energy expenditure while being transported \citep{bagnold1962auto}. This paper focuses on auto-suspension processes of the TCs rather than re-suspension (erosion), that is, we explore how the transported particles remain in the suspension state.

The auto-suspension mechanism of tiny particles was firstly proposed by \cite{knapp1938energy} and \cite{bagnold1962auto}. \cite{bagnold1962auto} pointed out that the particles could remain suspended when the vertical component of bed slope velocity $u_p^{//}$ is greater than the vertical sinking velocity of the particles  $w_p$, i.e. $u_p^{//}\sin\theta>w_p$, in which $\theta$ is the slope angle. With the auto-suspension of the TC and horizontal pressure difference, the flow will become more inclined to enter the self-acceleration mode \citep{pantin2011improved}. Then, the TC will erode the bottom bed and entrain the deposited sediments continuously. Because of the decrease of the horizontal pressure difference, the sedimentary bed can no longer be eroded or the terrain change. At the same time, the self-acceleration behaviour or ignition behaviour \citep{parker1982conditions,parker1986self} may be restricted or return to auto-suspension mode. In essence, auto-suspension is a prerequisite and necessary condition for self-acceleration \citep{parker1982conditions}.

Since the TC in the field is quite strong \citep{meiburg2015modeling}, it is very difficult to investigate auto-suspension by measuring the currents directly with instruments \citep{liu2012cyclone,xu2002distribution,xu2004situ}. Fortunately, after TCs, visible on-site traces, such as the source of sediments and the changes in riverbed elevation, provide convincing indirect evidence for the long-distance transport \citep{li2018gradual,andrieux2013turbidity}, which also proves the existence and importance of the auto-suspension of the TC. 

Previous laboratory and numerical studies had focused more on the hydrodynamic characteristics (such as Froude number), fluid velocity profile, particle concentration and sediment sequence to explain auto-suspension in the TC \citep{southard1981experimental,parker1987experiments,pantin2001experimental,pantin2011improved,sequeiros2009experimental,sequeiros2018internal}. \cite{pantin2001experimental} used the method of continuous inflow to conduct gravity flow experiments in a tubular channel on a steep slope ($\tan\theta=0.36$). He obtained that the evidence of auto-suspension is the sharp erosional upper bed surface, the increasing sediment proportion and the accelerated flow velocity. \cite{pantin2011improved} improved the experimental technique of auto-suspension in the laboratory on the basis of \cite{pantin2001experimental}. In both of the above studies, the interaction between the TC and the ambient fluid was avoided by a closed circular tube, which was different from the actual TC process. In contrast, the experiments of \cite{sequeiros2009experimental} and \cite{sequeiros2018internal} allowed TC to entrain the ambient water during the propagation. \cite{sequeiros2009experimental} used lightweight plastic particles to achieve auto-suspension on a gentle slope ($\tan\theta=0.05$) and pointed out that the finer bed material was easier to entrain, aiding the self-acceleration in the TC process. \cite{sequeiros2018internal} further revealed the positive feedback mechanism between the bed sediment entrainment and the flow velocity. And they proposed that the supercritical flow regime was a necessary and insufficient condition for self-acceleration and the thinning and fining rates along the turbidite could be adopted as an indicator to identify the acceleration of turbidity current. 

The Eulerian-Eulerian model is widely used to simulate the evolution of TCs, including the Reynolds-averaged Navier-Stokes (RANS) model, large-eddy simulation (LES) model and direct numerical simulation (DNS) model. They are all capable of giving general features of the TCs, however, RANS lacks an adequate description of the turbulent structures. The advantage of DNS is that it can provide more accurate results prior to the other two methods, but it requires a lot of computational resources. By comparison, LES gives relatively satisfactory results with less computational cost, which makes it widely adopted \citep{kyrousi2018large,goodarzi2020large,koohandaz2020prediction, pelmard2018grid}. In recent years, the coupled model of solving individual particle motion with the discrete element method (DEM) and fluid motion has been widely used to simulate two-phase flows, such as \cite{sun2016cfd,schmeeckle2014numerical,jing2016extended,maurin2018revisiting,pahtz2018cessation,pahtz2018universal,pahtz2020unification,zhu2020interface}. This approach does not evolve the Boussinesq approximation, different from the Eulerian-Eulerian model \citep{he2018investigations}. For TCs, \cite{biegert2017high} coupled a two-fluid model with a particle resolved model and  explored how a TC travelling along the surface of the sediment bed propagates within the bed. \cite{wildt2021two} employed an Eulerian–Lagrangian two-way coupled LES to simulate sediment plume development. \cite{xie2022fluid} employed the analysis of the dispersed TC particle movement and particle acceleration/deceleration to understand the evolution of the TC. However, such particle-based research is currently rare, and the effects of the complex interaction between dispersed particles and fluids in TC are still far from being understood, which is beneficial for the understanding of auto-suspension.

%Xie et al. (2021) used the LES-DEM model to study the lock-exchange TC on a flat slope. They distinguished the two stages of particle accelerating and decelerating and then exhibited the lift force were the key to TC evolution.

In order to investigate auto-suspension from the perspective of fluid-particle interactions and the movement of individual particles, this paper employs the LES-DEM model to simulate lock-exchange TCs on inclined slopes. In comparison with our earlier study on TC evolution over a flat slope \citep{xie2022fluid}, this present work focuses on the particle auto-suspension regimes and investigates the effects of different particle concentrations and slope angles on the auto-suspension regimes. We now additionally analyse the trajectories and spatial distribution of auto-suspension particles, the spatial characteristics of the forces acting on the auto-suspension particles and the auto-suspension index.
The remainder of the paper is as follows. Section \ref{sec:Description of numerical model} introduces the governing equations of the LES-DEM model, the fluid-particle interaction forces and the numerical settings. In Section \ref{sec:Mechanical explanation of particle auto-suspension}, the fluid and particle regimes of TC, the auto-suspension mechanism by particle statistics and fluid-particle interaction regimes, the energy budget and the auto-suspension criterion are demonstrated. Summary and conclusions are drawn in Section \ref{sec:Summary and conclusions}.

\section{\label{sec:Description of numerical model}Methodology}

In the LES-DEM approach, an Eulerian LES based on the open source code \href{https://openfoam.org/}{OpenFOAM} is adopted for a fluid phase coarser than the particle size to predict the dynamic flow process, and the DEM based on the \href{https://www.cfdem.com/}{LIGGGHTS} is adopted for the dispersed particle phase to accurately track the individual particles. These two models proceed independently, while the coupling of them at certain time intervals is implemented through the momentum exchange via CFDEMcoupling$^\circledR$, developed by \cite{kloss2012models}. This coupled LES-DEM model has been widely validated and applied \citep{wang2022super, yang2018numerical, blais2016development}, especially for TC simulations in our previous work \citep{xie2022fluid}. In the following, the equations for the fluid phase are described in subsection \ref{subsec:Governing equations for fluid phase}, and the equations for the particle phase and the fluid-particle interaction forces are given in subsection \ref{subsec:Discrete element method}. Subsection \ref{subsec:Numerical setup} elaborates on the numerical conditions of TC cases performed in the present work.

\subsection{\label{subsec:Governing equations for fluid phase}Governing equations for fluid phase}
The behaviour of the fluid phase is modelled by using LES, which is governed by the Navier-Stokes equations as follows \citep{chu2009cfd}:
\begin{equation}
\frac{\partial \alpha_f}{\partial t}+\frac{\partial \alpha _f \widetilde{u_{i,f}}}{\partial x_i} = 0,
\label{eq:1}
\end{equation}
\begin{equation}
\frac{\partial \alpha _f \rho _f \widetilde{u_{i,f}}}{\partial t}+
\frac{\partial \alpha _f \rho _f \widetilde{u_{i,f}} \widetilde{u_{j,f}}}{\partial x_j} =
-\frac{\partial \widetilde{p}}{\partial x_i} +
\alpha_f \rho_f g_i +
\frac{\partial \alpha _f \widetilde{\tau_{i j}}}{\partial x_j} -
R_{i,p f} +
\frac{\partial \alpha _f \widetilde{\Gamma_{i j}}}{\partial x_j},
\label{eq:2}
\end{equation}
where $\alpha_f$ is the fluid volume fraction in each computational cell ($\alpha_f = 1 - \alpha_p$), $\alpha_p$ is the particle volume fraction and $\widetilde{u_{i,f}}$, $\widetilde{p}$, $\widetilde{\tau_{i j}}$, $\widetilde{\Gamma_{i j}}$ are the filtered variables of fluid velocity, fluid pressure, fluid stress tensor and sub-grid stress tensor of the fluid phase, respectively; $\rho_f$ is the fluid density, and $g_i$ is the gravitational acceleration component; $R_{i, p f}$ is the momentum exchange with the particle phase, which is calculated by $R_{pf}=\sum\limits_{\xi=1}^{k_c}\mathbf{F}^f_\xi/V_{cell}$
with $\mathbf{F}^f$ being the fluid-particle interaction forces, $k_c$ the quantity of particles contained in the corresponding fluid cell and $V_{cell}$ the volume of a computational fluid cell. The Smagorinsky model, which has been widely utilized to model two-phase Eulerian--Lagrangian flows \citep{schmeeckle2014numerical,gui2018fine,elghannay2018simulations}, is adopted to resolve the sub-grid-scale (SGS) stress tensor as follows:
\begin{equation}
    \widetilde{\Gamma_{i j}} = 
    \rho_f 
    \left( \widetilde{u_{i,f}} \widetilde{u_{j,f}} - \widetilde{u_{i,f} u_{j,f}}
    \right) = 
    2 \mu_t \widetilde{S_{i j}} -
    \frac{1}{3} \widetilde{\Gamma _{k k}} \delta_{i j},
    \label{eq:3}
\end{equation}
where $\widetilde{S_{i j}}= (\partial \widetilde{u_{i,f}} / \partial x_j + \partial \widetilde{u_{j,f}} / \partial x_i)/2$, $\delta _{i j}$ is the Kronecker delta, and the SGS viscosity is obtained by
\begin{equation}
    \mu_t = \rho _f \left(C_s \Delta \right) ^2 \sqrt{2 \widetilde{S_{i j}} \widetilde{S_{i j}}},
    \label{eq:4}
\end{equation}
where $\Delta = (V_{cell})^{1/3}$ is the characteristic length scale and $C_s=0.1$ is a constant. The effect of the constant $C_s$ on the simulation results is investigated in Section \ref{subsec:Numerical setup}.

\subsection{\label{subsec:Discrete element method}Discrete element method}
The DEM is adopted to predict the track of each particle from the Lagrangian perspective. The governing equations for the translational and rotational motions of particle $i$ are based on Newton’s second law and the conservation law of angular momentum, which are calculated by \citep{jing2016extended,schmeeckle2014numerical}:
\begin{eqnarray}
m_i \frac{d\mathbf{u}_{p,i}}{d t} =
\mathbf{G}_i + \mathbf{F}^f_i + \sum\limits_{j = 1}^{n^c_i}\mathbf{F}^c_{i j},
\label{eq:5}
\end{eqnarray}
\begin{eqnarray}
I_i \frac{d\mathbf{\omega}_{p,i}}{d t} = \sum\limits_{j = 1}^{n^c_i}\mathbf{M}^c_{i j},
\label{eq:6}
\end{eqnarray}
where $m_i$ and $I_i$ are the mass and the moment of inertia of particle $i$, respectively, $\mathbf{u}_{p,i}$ and $\mathbf{\omega}_{p,i}$ are the translational velocity and angular velocity of particle $i$, respectively, $n^c_i$ represents the number of contacting particles around particle $i$, $\mathbf{F}^c_{i j}$ and $\mathbf{M}^c_{i j}$ are, respectively, the contact force and the torque acting on the particle $i$ by particle $j$ or the boundary wall, $\mathbf{F}^f_i$ denotes the fluid-particle interaction force acting on the particle $i$ and $\mathbf{G}_i = m_i \mathbf{g}$ is the gravity of particle $i$. Note that fluid-induced torque is not involved in Eq. (\ref{eq:6}). This is because the fluid cell is always several times larger than the diameter of the particle, and the fluid variables are locally averaged over the fluid cell \citep{jing2016extended}.

In the research, the fluid-particle interaction force $\mathbf{F}^f$ comprises the buoyancy $\mathbf{F}^b$, drag force $\mathbf{F}^d$, lift force $\mathbf{F}^l$ and added mass force $\mathbf{F}^{add}$, i.e., $\mathbf{F}^f=\mathbf{F}^b+\mathbf{F}^d+\mathbf{F}^l+\mathbf{F}^{add}$. The buoyancy $\mathbf{F}^b$ acting on a single particle with diameter $d_p$ is calculated by
\begin{equation}
    \mathbf{F}^b=-\frac{1}{6}\upi\rho_f d_p ^3 \mathbf{g},
    \label{eq:7}
\end{equation}
with gravitational acceleration $\mathbf{g} = [0,0,-9.81] \mathrm{m/s^2}$.

The drag force $\mathbf{F}^d$ is expressed as follows \citep{di1994voidage}:
\begin{equation}
    \mathbf{F}^d=\frac{1}{8} C_D \rho _f \upi d_p ^2 \left(
    \mathbf{u}_f - \mathbf{u}_p
    \right)|\mathbf{u}_f - \mathbf{u}_p| \alpha _f ^{-\chi},
    \label{eq:8}
\end{equation}
where $C_D$ is the drag coefficient and $\chi$ is the correction factor, which are respectively given by
\begin{eqnarray}
C_D = \left(0.63+\frac{4.8}{\sqrt{\alpha _f \Rey_p}}
\right)^2,
\label{eq:9}
\end{eqnarray}
\begin{eqnarray}
\chi = 3.7 -0.65 \mathrm{exp}\left[
-\frac{\left(
1.5-\mathrm{log}_{10}\left(\alpha_f \Rey_p
\right)
\right)^2}{2}
\right],
\label{eq:10}
\end{eqnarray}
where $\Rey_p$ is the particle Reynolds number that can be expressed as
\begin{equation}
    \Rey_p = \frac{\rho_f d_p |\mathbf{u}_f - \mathbf{u}_p|}{\mu _f}
    \label{eq:11}.
\end{equation}

The lift force $\mathbf{F}^l$ here follows \cite{loth2009equation}:
\begin{equation}
    \mathbf{F}^l=\frac{1}{8} C_L \rho _f \upi d_p ^2 |\mathbf{u}_f - \mathbf{u}_p|
    \left[
    \left(
    \mathbf{u}_f - \mathbf{u}_p
    \right)
    \times \frac{\mathbf{\omega}_f}{|\mathbf{\omega}_f|}
    \right],
    \label{eq:12}
\end{equation}
where $\mathbf{\omega}_f$ is the fluid vorticity, and $C_L$ is the lift coefficient given by \citep{mclaughlin1991inertial,loth2009equation}:
\begin{equation}
    C_L=J^{\ast}\frac{12.92}{\upi}\sqrt{\frac{\omega^\ast}{\Rey_p}} +
    \Omega^\ast_{p,e q} C^\ast_{L,\Omega},
    \label{eq:13}
\end{equation}
with $\omega^\ast=|\mathbf{\omega}_f|d_p / |\mathbf{u}_f - \mathbf{u}_p|$. The function  $J^\ast$ in (\ref{eq:13}) reads \citep{mei1992approximate}:
\begin{equation}
    J^\ast = 0.3 \left\{
    1 + \tanh \left[ 
    \frac{5}{2} \left(
    \mathrm{log}_{10}\sqrt{\frac{\omega^\ast}{\Rey_p}} + 0.191
    \right)
    \right]
    \right\}
    \left\{
    \frac{2}{3} + \tanh\left[ 6 \sqrt{\frac{\omega^\ast}{\Rey_p}}\right]
    \right\}.
    \label{eq:14}
\end{equation}
Furthermore, the empirical correction $C^\ast_{L,\Omega}$ and empirical model for $\Omega^\ast_{p,eq}$ in (\ref{eq:13}) are given by \citep{loth2009equation}:
\begin{equation}
    \Omega^\ast_{p,e q}=\frac{\omega^\ast}{2}\left(
    1-0.0075\Rey_\omega
    \right)\left(
    1-0.062\sqrt{\Rey_p} - 0.001\Rey_p
    \right),
\label{eq:15}
\end{equation}
\begin{equation}
    C^\ast_{L,\Omega} = 1 - \left\{
    0.675 + 0.15\left(
    1+\tanh\left[0.28 \left(\Omega^\ast_p-2
    \right)
    \right]
    \right)
    \right\}
    \tanh\left[0.18 \sqrt{\Rey_p}
    \right],
\label{eq:16}
\end{equation}
where $\Rey_\omega=\rho_f |\mathbf{\omega}_f|d_p ^2 / \mu _f$ and $\Omega^\ast_p = |\mathbf{\omega}_p|d_p/|\mathbf{u}_f - \mathbf{u}_p|$.

The added mass force $\mathbf{F}^{a d d}$ is formulated by
\begin{equation}
    \mathbf{F}^{a d d} = \frac{1}{6}C_{a d d}\rho _f \upi d_p ^2 \left(
    \frac{D\mathbf{u}_f}{D t} - \frac{D\mathbf{u}_p}{D t}
    \right),
    \label{eq:17}
\end{equation}
where $C_{a d d}=0.5$ is the added mass coefficient. Although the added mass force on particles in open channel sediment transport studies is generally considered to be unimportant \citep{schmeeckle2014numerical,pahtz2021unified}, we found in our previous study \citep{xie2022fluid} that it cannot be neglected in TCs. Equation~(\ref{eq:17}) assumes that particles are not in close proximity to one another. Due to the low particle volume fraction, this is approximately the case in our TC simulations. Other forces, such as the Basset force, are always secondary (several orders of magnitude smaller) as compared with these predominant forces, and are difficult to quantify analytically \citep{duran2011aeolian, schmeeckle2014numerical}.%, but its performance, during the evolution of the TC on the inclined slope where the accelerations of the two phases exert uncertainty, deserves to be re-evaluated.

In coupling two phases, the fluid-particle interactions are transmitted only by the momentum exchange term ($R_{i,pf}$ in (\ref{eq:2})). The effect of particles on SGS stresses of the fluid is ignored due to low particle Reynolds number ($\Rey=0.01\sim 0.1$) and small particle--turbulence length-scale ratio, $d_p / \eta \ll 1$, where $\eta \sim O (10^{-3})$ is the Kolmogorov scale  \citep{bagchi2004response,elghobashi1992direct}. This means small vortical structures caused by individual particles are negligible. In addition, the turbulence modulates the flow by the SGS model, which then impacts the particle movement, and we do not model the effect of the sub-grid fluid field on the particles, since this effect is relatively weak when the LES velocity field is well resolved \citep{armenio1999effect}.

Assuming that the particles are soft spheres, the particle--particle contact force $\mathbf{F}^c_{ij}$ can be evaluated by using an elastic spring and a viscous dashpot \citep{cundall1979discrete}. The contact force caused by the collision, which includes the normal component $\mathbf{F}^n_{ij}$ and tangential component $\mathbf{F}^t_{ij}$, is written as
\begin{equation}
    \begin{array}{l}
{\bf{F}}_{ij}^c = {\bf{F}}_{ij}^n + {\bf{F}}_{ij}^t\\
 = \left\{ {\begin{array}{*{20}{c}}
{\left( {{k^n}\delta _{ij}^n - {\gamma ^n}\nu _{ij}^n} \right) + \left( {{k^t}\delta _{ij}^t - {\gamma ^t}\nu _{ij}^t} \right),}&{{\bf{F}}_{ij}^t < {\mu _c}{\bf{F}}_{ij}^n}\\
{\left( {{k^n}\delta _{ij}^n - {\gamma ^n}\nu _{ij}^n} \right) + {\mu _c}{\bf{F}}_{ij}^n,}&{{\bf{F}}_{ij}^t \ge {\mu _c}{\bf{F}}_{ij}^n}
    \end{array}} \right.
    \end{array}
\end{equation}
where $\delta^n_{ij}$, $\delta^t_{ij}$, $\nu^n_{ij}$ and $\nu^t_{ij}$ are the normal overlap distance, the tangential displacement, the normal relative velocity of particles $i$ and $j$ and the tangential relative velocity of particles $i$ and $j$, respectively. Here, $\mu_c$ ($=0.5$ adopted in this study) is the coefficient of friction, $k^n$, $k^t$, $\gamma^n$ and $\gamma^t$ are the elastic constants for normal contact and tangential contact and the viscoelastic damping constants for normal contact and tangential contact, respectively. These parameters are determined by a Young’s modulus $Y=5\times10^6$ Pa, Poisson ratio $\nu=0.45$ and coefficient of restitution $e=0.3$, which are the inherent properties of the particle material.

\subsection{\label{subsec:Numerical setup}Numerical set-up}
\begin{figure}
\centerline{\includegraphics[width=\textwidth]{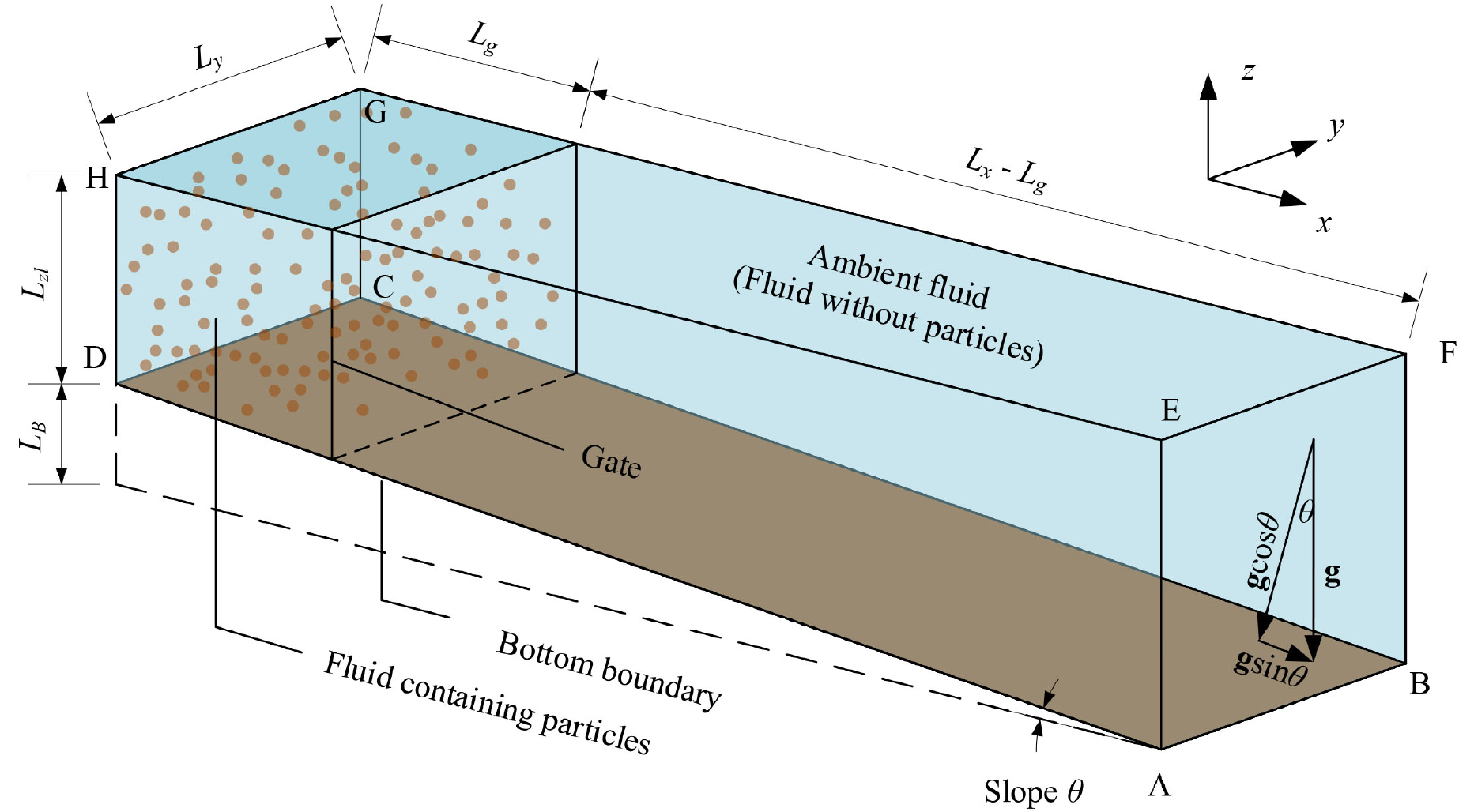}}% Here is how to import EPS art
\caption{\label{fig:Schematic view} Schematic view of the initial configuration of the lock-exchange TC.}
\end{figure}

The set-up of the numerical experiment is shown in figure \ref{fig:Schematic view}%
.
The numerical simulation domain is a hexahedral tank filled with water with a slope. The angle between the slope and the horizontal plane is defined as $\theta$ and $\tan\theta=L_B/L_x$, in which $L_B$ is the maximum lift height of the slope and $L_x$ represents the length of the hexahedral tank. Different boundary conditions are employed in our simulation. The bottom wall and the top wall in the vertical ($z$) direction are set as a no-slip wall boundary condition and a free-slip boundary condition, respectively. The longitudinal ($x$) walls (the left and right surfaces) and transverse ($y$) walls (the front and back surfaces) have no-slip wall boundary conditions and periodic boundary conditions, respectively. In figure \ref{fig:Schematic view}, the distance between the gate and the left wall is denoted by $L_g$, and the turbidity current on the left of the gate is filled with dispersed particles. At the beginning of the simulation, the gate is pulled away and the particles move forward during the settling process, thus forming a TC. The TC head refers to the foremost part of the current, the horizontal length of which is defined as 0.15 times the length of the TC. The definition of different head lengths ($0.10\sim0.20$ times the length of the current) does not substantially affect the quality of the results of the head average forces \citep{xie2022fluid}, however, its effect on the head auto-suspension index is unknown, which is explored in Section \ref{subsec:Criterion for auto-suspension}. 

In the following, all parameters are dimensionless (except some of the parameters of the particles). We take half of the left water depth ($L_{zl}=L_z - L_B$) in the domain as the characteristic length $L_{zl}/2$ and the buoyancy velocity $u_b$ as the characteristic velocity, which is calculated as follows:
\begin{equation}
    u_b=\sqrt{C_0|\mathbf{g}|\frac{\left(\rho_p - \rho_f
    \right)L_{z_l}}{2\rho_f}},
    \label{eq:19}
\end{equation}
where $C_0$ denotes the initial particle concentration. 
The dimensionless time is given by $2tu_b/L_{zl}$. The forces acting on particles are made dimensionless by the effective gravity $\mathbf{G}'$, which is the absolute value of the resultant force of buoyancy $\mathbf{F}^b$ and gravity $\mathbf{G}$ ($=\rho_p \upi d_p ^3 \mathbf{g} / 6$), that is $\mathbf{G}'=\mathbf{G}+\mathbf{F}^b = (\rho_p-\rho_f)\upi d_p^3 \mathbf{g}/6$.

As we have a slope in the TC case shown in figure \ref{fig:Schematic view}, we employ the variables $\psi$ in the bed-parallel and bed-normal directions ($\psi^{//}$ and $\psi^\bot$) for the convenience of analysis, which are calculated as follows:
\begin{equation}
\psi^{//} = \psi_x\cos\theta-\psi_z\sin\theta,
\label{eq:22}
\end{equation}
\begin{equation}
\psi^{\bot}= \psi_x\sin\theta+\psi_z\cos\theta,
\label{eq:23}
\end{equation}
with a longitudinal ($x$) variable $\psi_x$ and a vertical ($z$) variable $\psi_z$.

Numerical settings of different cases are shown in table~\ref{tab:case table}. 
In this study, the effects of different particle concentrations and different inclined slopes are of concern in cases 1,2,3 and cases 1,4,5,6, respectively. Fluid grid size for all coordinates is $2\sim4$ times the particle diameter $d_p$. The dimensional particle diameter is 0.00005 m (50 $\mu$m) and the particle density $\rho_p$ is 1200 $\mathrm{kg/m^3}$. The density of the fluid is 1000 $\mathrm{kg/m^3}$, thus the ratio of the densities $s=\rho_p/\rho_f=1.2$. The particle Reynolds number $\Rey_p$ is relatively small (about $0.01\sim0.1$), and the particle Stokes number $St=s \rho_f | \mathbf{u}_f - \mathbf{u}_p | d_p/(9\mu_t)$ is $O(10^{-3}\sim 10^{-1})$, indicating that the particle flow ought to follow the fluid streamlines of the TCs and not be determined by the inertia of the particles. The front Reynolds number $\Rey_f$ is given by $\Rey_f = \rho_f u_{front} L_h / \mu_t$, where $u_{front}$ is the front velocity, $L_h$ is the height of the TC head and $\mu_t$ is the fluid dynamic viscosity. The Reynolds number $\Rey$ is defined by $\Rey=\rho_f u_b L_{zl}/(2\mu_t)$. The initial particle concentration $C_0$ varies from 1\% to 2\%. 

\begin{table}
\caption{\label{tab:case table}Parameters of numerical simulations for each case.}
  \begin{center}
\def~{\hphantom{0}}
\begin{tabular}{ccccccc}
 & Case 1 & Case 2 & Case 3 & Case 4 & Case 5 & Case 6\\ \hline
Domain dimensions & & & & & & \\
 $L_x \times L_y \times L_{z l}$ & \multicolumn{6}{c}{$10 \times 1 \times 2$}\\
 $L_z$ & 3 & 3 & 3 & 4 & 2.67 & 2.5\\
 $L_B$ & 1 & 1 & 1 & 2 & 0.67 & 0.5\\
 $L_g$ & 2 & 2 & 2 & 2 & 2 & 2\\
 $\tan\theta$ & 1/10 & 1/10 & 1/10 & 1/5 & 1/15 & 1/20\\ \hline
Mesh resolutions & & & & & & \\ 
 $N_x$ & 250 & 250 & 250 & 250 & 250 & 250\\
 $N_y$ & 25 & 25 & 25 & 25 & 25 & 25\\
 $N_z$ & 80 & 80 & 80 & 100 & 80 & 80\\ \hline
Other properties & & & & & & \\
 $C_0$ & 0.010 & 0.015 & 0.020 & 0.010 & 0.010 & 0.010\\
 $\Rey_f$  & 21$\sim$53 & 21$\sim$62 & 21$\sim$69 & 32$\sim$58 & 17$\sim$50 & 14$\sim$48\\
 $\Rey $  & 50 & 61 & 70 & 50 & 50 & 50\\
\end{tabular}
  \end{center}
\end{table}

\begin{figure}
\centerline{\includegraphics[width=\textwidth]{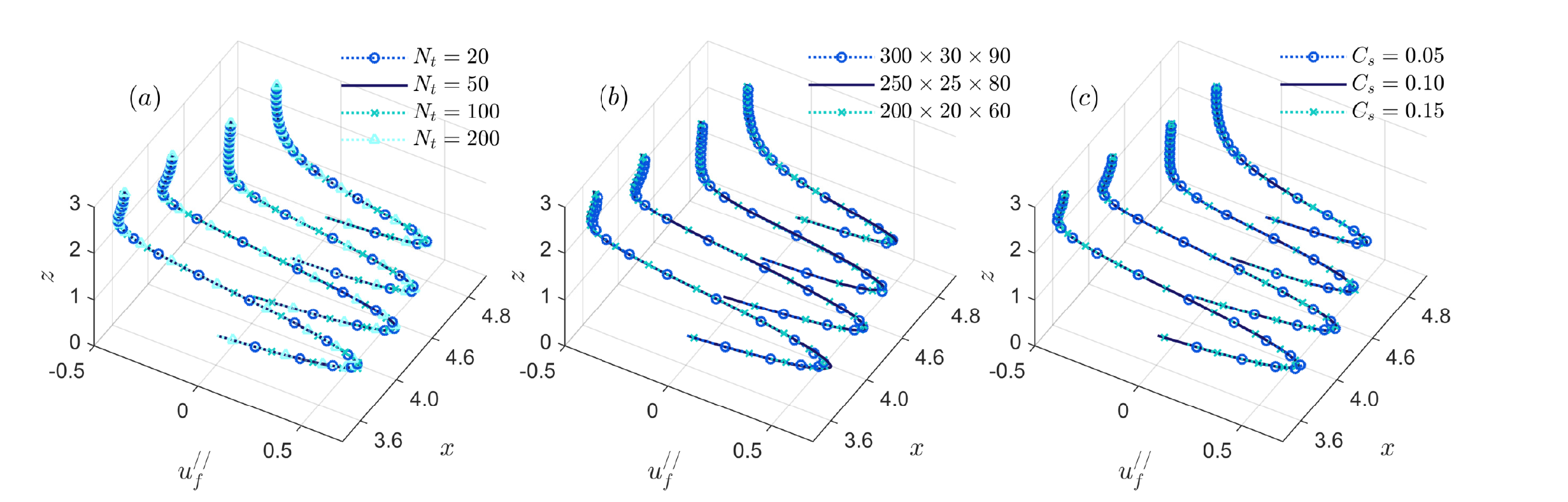}}% Here is how to import EPS art
\caption{\label{fig:mesh sensitivity profile} Fluid velocity profiles at the four selected positions ($x$ = 3.6, 4.0, 4.6 and 4.8) at $t=6$ for case 1 ($a$) with four different coupling intervals ($N_t=$ 20, 50, 100 and 200), ($b$) with three different computational grid resolutions ($300(N_x)\times30(N_y)\times90(N_z)$, $250\times25\times80$ and $200\times20\times60$) and ($c$) with three different Smagorinsky constants ($C_s=$ 0.05, 0.10 and 0.15).}
\end{figure}

The DEM time step is set to $10^{-7}$ s, which is lower than the Rayleigh critical time step proposed by \cite{li2005comparison}. To improve computational efficiency, we set the fluid-particle coupling interval $N_t$ to be 100. In other words, the momentum exchange between the particle and fluid phases is integrated every 100 DEM time steps. At this time, the CFD time step is equal to $10^{-5}$ s, which corresponds to a Courant number less than 0.01. Here, we briefly discuss the effect of the coupling interval $N_t$ on the simulation results. We keep the DEM time step unchanged and adjust the coupling intervals to 20, 50, 100 and 200. The comparisons of the transverse-averaged fluid velocity at four different positions  ($x$ = 3.6, 4.0, 4.6 and 4.8) at $t=6$ for case 1 with four coupling intervals $N_t$ (20, 50, 100, and 200) are shown in figure \ref{fig:mesh sensitivity profile}($a$). The profiles for various coupling intervals $N_t$ are very similar, indicating that the model results are not sensitive to $N_t$. It is worth mentioning that an excessively large coupling interval may cause particles to move in multiple grids within one CFD time step, which will lead to uncertainty in interphase momentum exchange.

The accuracy of the LES model is affected by the quality of the grid resolution. \cite{pelmard2018grid} proposed that a good resolution can be obtained if the ratio of the SGS viscosity to the molecular $N_v$ in the low turbulence region above the upper boundary of the TC is lower than 0.3 and the ratio of the SGS shear stress to the resolved Reynolds stress $N_s$ inside the turbulent mixing layer satisfies $N_s<0.05$. When using the grid resolutions in the table \ref{tab:case table}, $N_v$ is of the order of $O(10^{-4}\sim 10^{-6})$ and $N_s$ in the order of $O(10^{-4})$ in TC simulations, which meet the requirements of \cite{pelmard2018grid}. This demonstrates that the grid resolutions are fine for LES. Moreover, the effect of different grid resolutions on the results is investigated. The comparisons of the transverse-averaged fluid velocity at four different positions with three meshes ($300 \times 30 \times 90$, $250\times 25\times 80$ and $200\times20\times60$) are shown in figure \ref{fig:mesh sensitivity profile}($b$). The velocity profiles essentially coincide with each other, which means the three grids give consistent results for the current and the numerical model set-up here is reliable. As shown in table \ref{tab:case table}, the grid of $250\times25\times80$ cells for all cases except case 4 is adopted in our simulations. For case 4, the slope angle is large and $L_z$ is 4 in table \ref{tab:case table} and the grid is $250\times 25\times 100$. 

In general, the constant $C_s$ in the Smagorinsky model needs to be adjusted for different flow events \citep{katopodes2018free}. We discuss the sensitivity of $C_s$ here. Figure \ref{fig:mesh sensitivity profile}($c$) shows the transverse-averaged fluid velocity at four positions with three constants $C_s$ (0.05, 0.10 and 0.15). As can be observed, the simulation of the TCs in this work is essentially unaffected by the changes in the constant $C_s$. A possible reason for this could be the fact that the particle Reynolds number is very low ($\Rey_p = 0.01\sim0.1$), suppressing vortices at the particle scale (note that also the flow Reynolds number is not particularly large, $O(10))$. This indicates that the details of the Smagorinsky model are not very relevant for our cases.

\section{\label{sec:Mechanical explanation of particle auto-suspension}Results and discussion}

\subsection{\label{subsec:Validation of front positions}Development of the TC}

\begin{figure}
\centerline{\includegraphics[width=\textwidth]{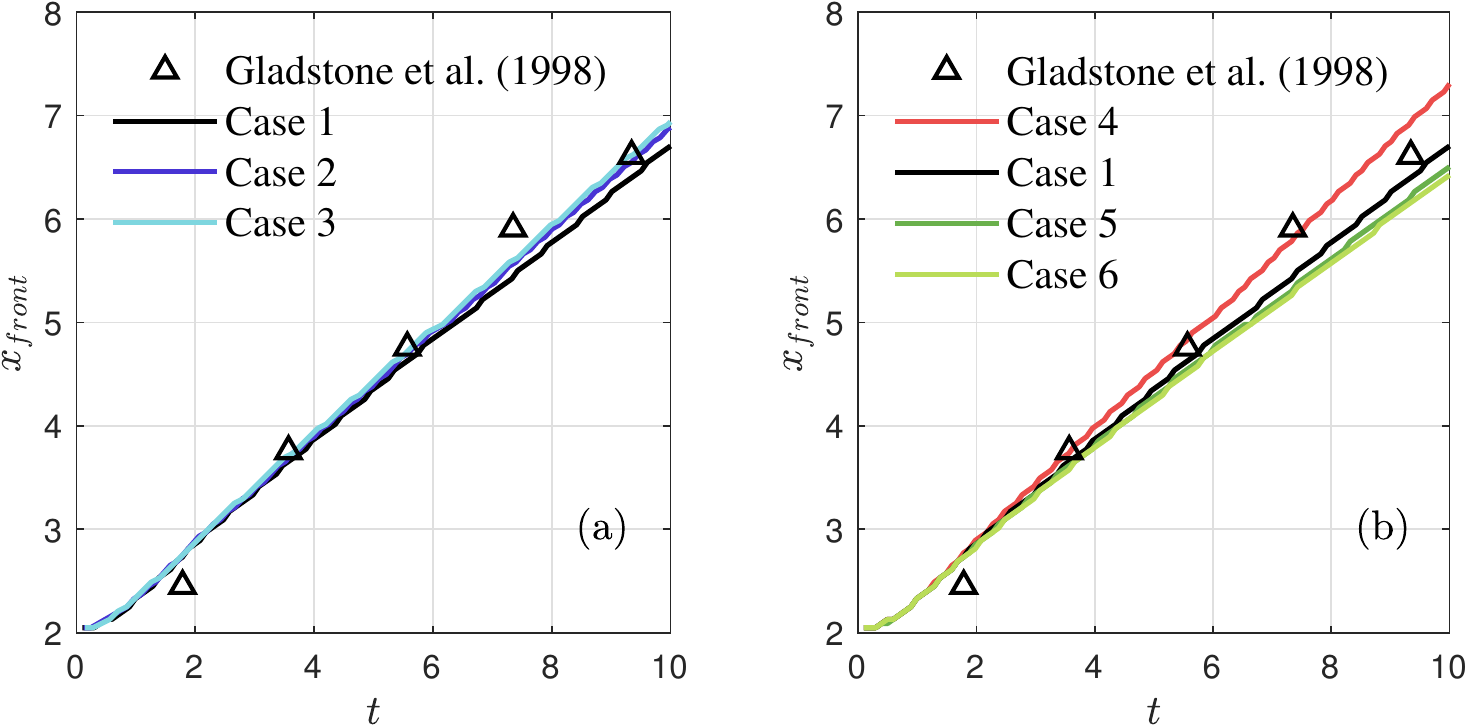}}% Here is how to import EPS art
\caption{\label{fig:head position} Temporal evolution of simulated TC fronts under different cases, and the comparison of simulated fronts with the measured data \citep{gladstone1998experiments}.}
\end{figure}

The TC front $x_{front}$, an important parameter in TC, is defined as the position of the particle moving furthest in the longitudinal direction. 
Figure \ref{fig:head position} plots the comparison of numerical and experimental results for the time evolution of the front position. The simulation results are consistent with the experimental data by \cite{gladstone1998experiments}, which shows the feasibility of the model for simulating TCs. It can be seen from the gradient of the curves (the front velocity $u_{front}$) in figures \ref{fig:head position}(\textit{a}) and \ref{fig:head position}(\textit{b}) that the front velocity reaches its maximum rapidly and then remains roughly unchanged. The dimensionless velocities increase slightly with the increase of the particle concentration in figure \ref{fig:head position}(\textit{a}). The characteristic velocities $u_b$ are 0.0099 m/s, 0.0121 m/s and 0.0140 m/s for cases 1, 2 and 3, respectively. This means that the actual front velocity $u_{front}$ increases positively in relation to the increase in initial particle concentration, which is consistent with the understanding of previous studies \citep{hu2020layer,bonnecaze1993particle}. In figure \ref{fig:head position}(\textit{b}), as the slope angle increases ($\theta<20^\circ$), the gradient becomes larger, indicating a larger front velocity of the TC \citep{he2018investigations,steenhauer2017dynamics}. The steepening of the slope raises the current barycentre, resulting in an increasing available particle potential energy, which will be converted into the kinetic energy of the system.

%\subsection{\label{subsec:Fluid velocity field and vorticity field}Fluid velocity profile}

In previous studies, the fluid velocity profile of the TC is generally similar \citep{altinakar1996flow}. The TC area can be divided into two regions, the wall region and the jet region, which is based on the location of the maximum fluid velocity parallel to the slope $u_f^{//,\max}$. The main control factors of the two regions are different: the wall region (lower layer) is dominated by the bottom wall friction \citep{nourmohammadi2011experimental}, whereas the jet region (upper layer) is dominated by the ambient fluid shear. The fluid velocity distribution of the jet region and the wall region can be described as fitted functions \citep{altinakar1996flow,nourmohammadi2011experimental}:
\begin{eqnarray}
    \frac{u_f  ^{//} \left( Z \right)}{u_f^{//, \max}} =\left\{
    \begin{array}{cc}
    \mathrm{exp} \left[ - \beta\left(
    \frac{Z-H_m}{H-H_m}
    \right)^\gamma
    \right], & \mathrm{the\;jet\; region} \\
    \left(
    \frac{Z}{H_m}
    \right)^{1/n}, & \mathrm{the\;wall\; region}
\end{array}\right.
    \label{eq:20}
\end{eqnarray}
where $u_f^{//} (Z)$ is the fluid bed-parallel velocity at a distance of $Z$ from the slope, $H_m$ is the distance between the location of $u_f^{//,\max}$ and the slope and $H$ denotes the depth-averaged current height which is expressed as follows:
\begin{equation}
    H=\frac{\left(\int\limits_0^{Z_{top}} 
    u_f^{//} d Z
    \right)^2}{\int\limits_0^{Z_{top}} \left(
    u_f^{//}
    \right)^2 d Z},
    \label{eq:21}
\end{equation}
where $Z_{top}$ is the farthest distance from the slope for which the fluid bed-parallel velocity is positive. Here, $\beta$, $\gamma$ and $n$ are empirical coefficients. %
%\begin{figure}
%\centerline{\includegraphics[width=3.3in]{Figure3.pdf}}% Here is how to import EPS art
%\caption{\label{fig:validation profile} Fluid velocity profiles at the four selected positions ($x$ = 3.0, 3.6, 4.2, and 4.8) at $t=9.90$ for Case 1, and the fitted results for profile data.}
%\end{figure} 
We now proceed to employ (\ref{eq:20}) to fitted fluid velocity profiles at four different locations ($x$ = 3.0, 3.6, 4.2 and 4.8) at $t=10$. Table \ref{tab:fitting result table} shows the tuned parameters and the fitting results for all cases. The simulated fluid velocity profiles can be well fitted with (\ref{eq:20}) (the determination coefficients $\mathrm{R}^2$ are all higher than 0.95). It proves the reasonable performance of the model for reproducing TC processes.

\begin{table}
\caption{\label{tab:fitting result table} Comparison of fitting results for present six cases.}
  \begin{center}
\def~{\hphantom{0}}
\begin{tabular}{cccccc}
 & \multicolumn{2}{c}{Wall region} & \multicolumn{3}{c}{Jet region} \\
 \cline{2-6}
 & $n$ & $\mathrm{R}^2$ & $\beta$ & $\gamma$ & $\mathrm{R}^2$\\
 \hline
Case 1 & 1.67 & 0.95 & 0.63 & 2.68 & 0.99\\
Case 2 & 1.77 & 0.95 & 0.73 & 2.62 & 0.99\\
Case 3 & 1.74 & 0.95 & 0.75 & 2.62 & 0.99\\
Case 4 & 1.77 & 0.95 & 0.77 & 2.61 & 0.99\\
Case 5 & 1.63 & 0.95 & 0.59 & 2.64 & 0.99\\
Case 6 & 1.60 & 0.95 & 0.56 & 2.64 & 0.99\\
\end{tabular}
%\end{ruledtabular}
\end{center}
\end{table}

\begin{figure}
\centerline{\includegraphics[width=\textwidth]{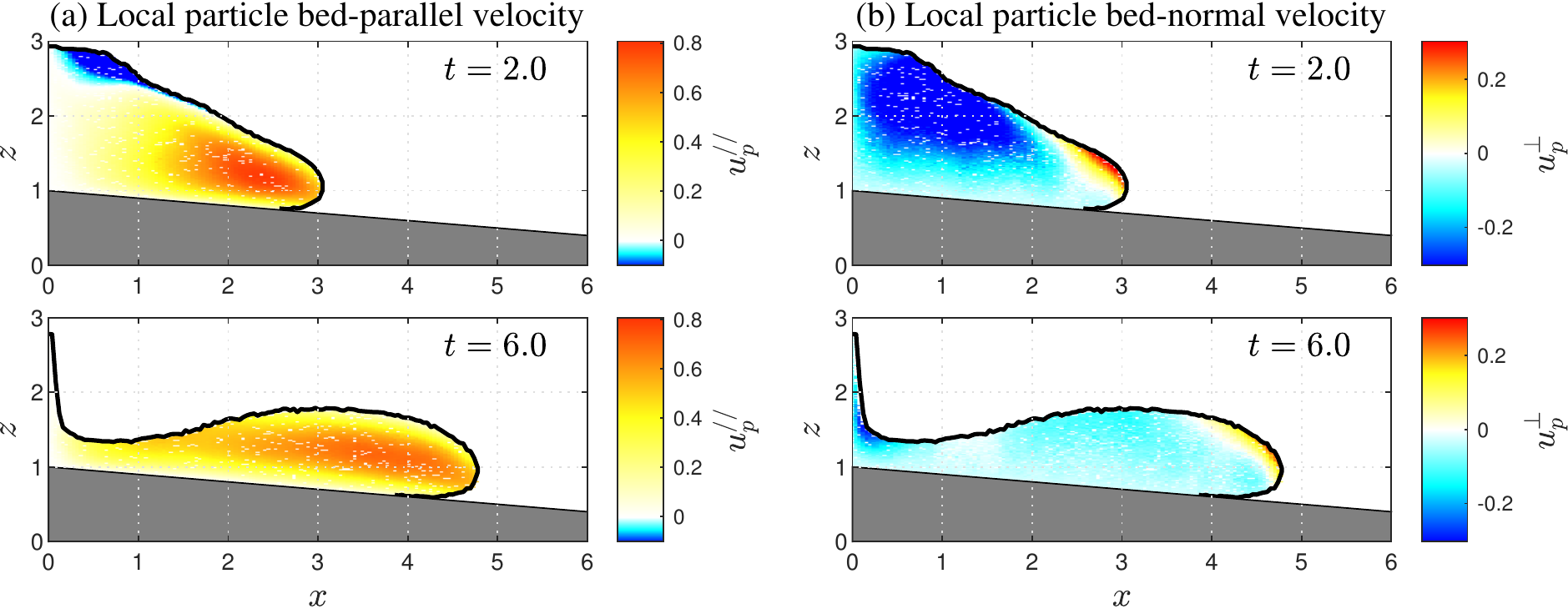}}% Here is how to import EPS art
\caption{\label{fig:profile} Spatial variations of (\textit{a}) local average particle bed-parallel velocity and (\textit{b}) local average particle bed-normal velocity at two selected times. Black line indicates the TC profile.}
\end{figure}

The local average spatial variations of the bed-parallel ($u^{//}_{p,i}$) and bed-normal ($u^\bot_{p,i}$) velocities at $t=2$ and $t=6$ in case 1 are shown in figure \ref{fig:profile}
. 
In figure \ref{fig:profile}(\textit{a}), the current is mainly transported downstream along the slope in the TC except for the particles in the top layers at the beginning stage \citep{hitomi2021measurement,wildt2020cfd,nourmohammadi2011experimental}. The reason why top particles are transported upstream (blue coloured) is the intrusion flow. Then all particles move downstream at $t=6$. For the bed-normal velocities shown in figure \ref{fig:profile}(\textit{b}), most particles are gradually settling to the bottom wall due to the dominance of gravity, while $u^\bot_{p,i}$ of the particles in the TC head border region is positive or zero on average. This exhibits that they cannot only stay suspended, but can even rise beyond the suspension state (defining them as auto-suspension particles or highly auto-suspension particles). 

\begin{figure}
\centerline{\includegraphics[width=\textwidth]{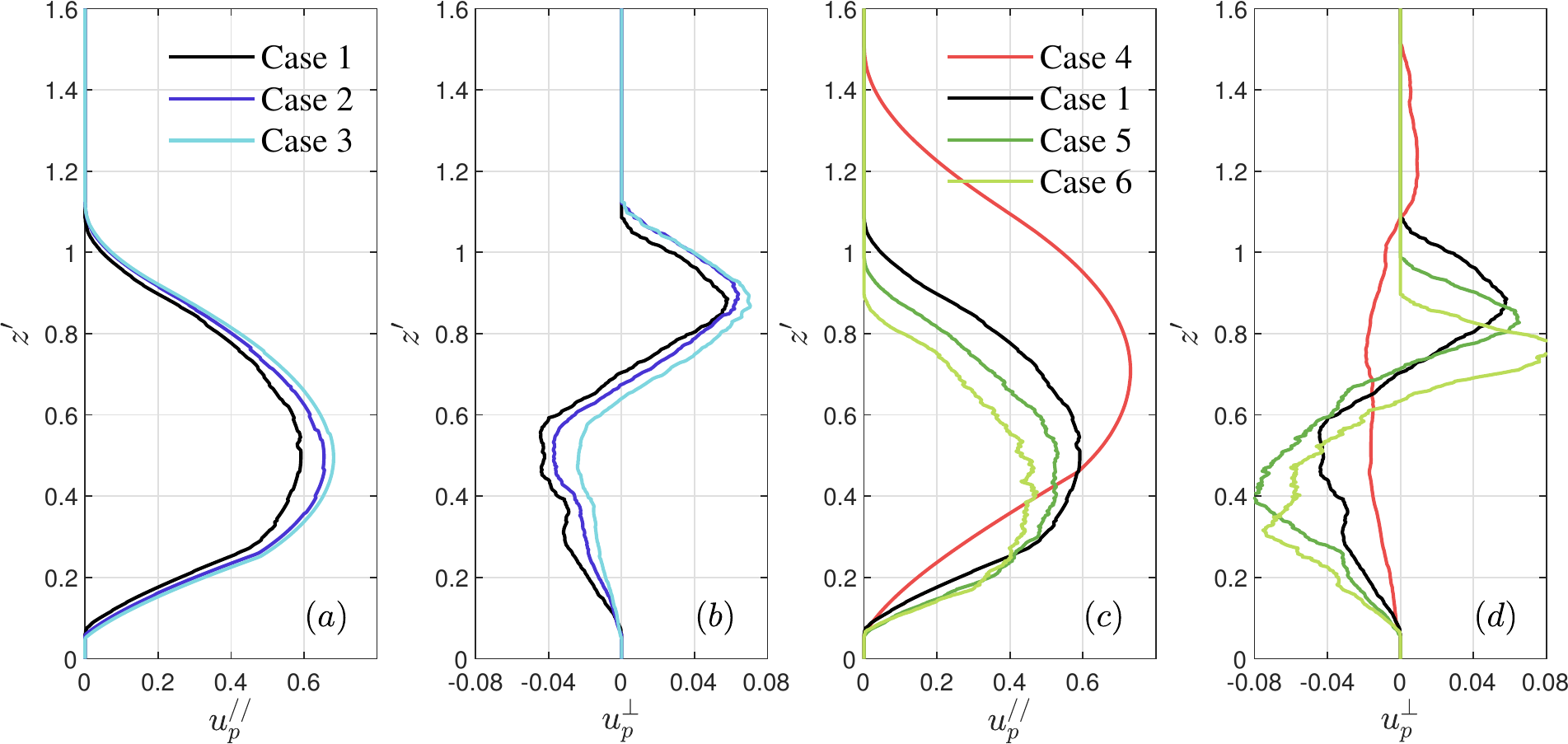}}% Here is how to import EPS art
\caption{\label{fig:particleVertical} ($a,c$) Particle bed-parallel velocity profiles and ($b,d$) particle bed-normal velocity profiles at 0.05 current lengths upstream from the front at $t=6$ under different cases. Here, $z'=z-z_b$ with $z_b$ denoting the bed elevation. $\Rey_f$ (case 1) = 27, $\Rey_f$ (case 2) = 35, $\Rey_f$ (case 3) = 46, $\Rey_f$ (case 4) = 37, $\Rey_f$ (case 5) = 24, and $\Rey_f$ (case 6) = 23.}
\end{figure}

Given that particle auto-suspension occurs mostly in the head, we display the profiles of particle bed-parallel and bed-normal velocities at 0.05 current lengths upstream from the front at $t=6$ under different cases, as shown in figure \ref{fig:particleVertical}. 
For different particle concentrations at the same slope angle, the front positions shown in figure \ref{fig:head position} are quite similar to cases 1-3. As shown in figure \ref{fig:particleVertical}($a$), with increasing particle concentration, the particle bed-parallel velocity increases, which agrees well with the previous study \citep{hu2020layer}. The different directions of bed-normal velocity of particles in the upper and lower layers of the head exhibit the separation of particles in the TC head. In figure \ref{fig:particleVertical}($b$), it is observed that most particles in the lower layer move toward the slope while those in the upper layer ($z'>0.7$) move away from the slope, which can be directly viewed in the spatial variation of case 1 in figure \ref{fig:profile}($b$). In the lower layers, the fluids are subjected to wall viscous shear, resulting in a low fluid velocity, which is insufficient to achieve particle suspension, while the velocity of the fluids in the upper layers is strong enough to help achieve suspension. 
With the increase of the slope angle shown in figure \ref{fig:particleVertical}($c$), the maximum particle velocity along the slope increases, and the height of the TC grows, which is really different from figure \ref{fig:particleVertical}($a$). It manifests itself as an increase in the ability of auto-suspension of the TC head. Note that the increasing slope angle thickens the lower region (figure \ref{fig:particleVertical}($d$)) where the particle bed-normal velocity is negative, and it simultaneously inhibits the upward movement of the upper particles and the settling of the lower particles, implying the decrease in the separation rate of the upper and lower particles.

\begin{figure}
\begin{tabular}{c}
\centerline{\includegraphics[width=0.7\textwidth]{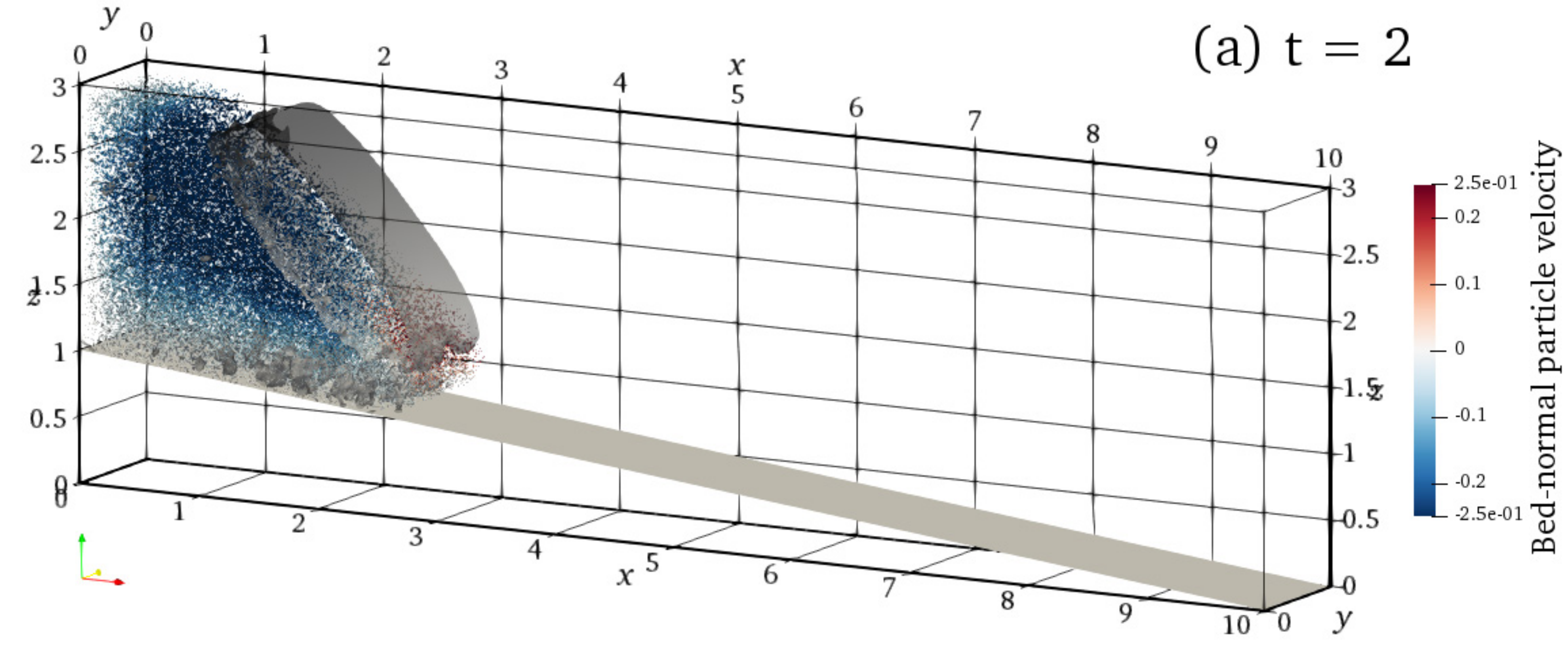}} \\
\centerline{\includegraphics[width=0.7\textwidth]{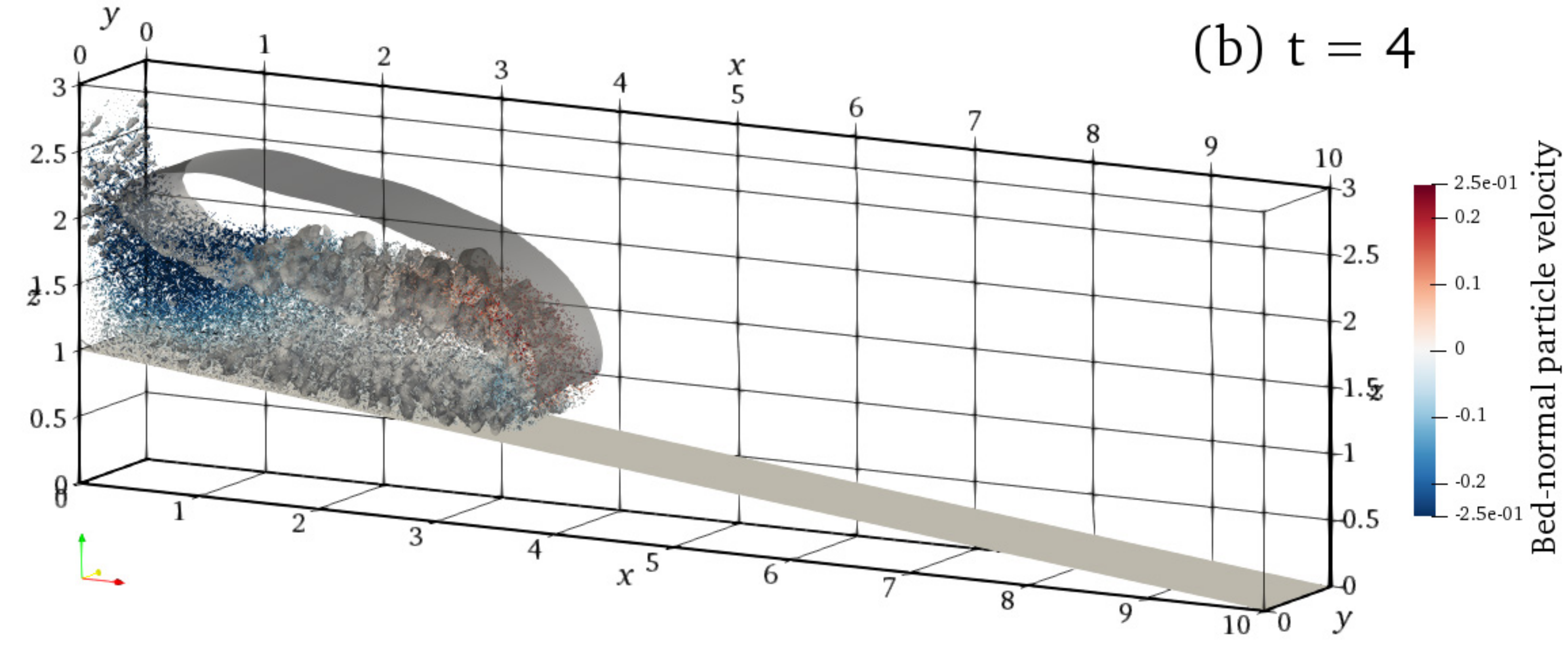}} \\
\centerline{\includegraphics[width=0.7\textwidth]{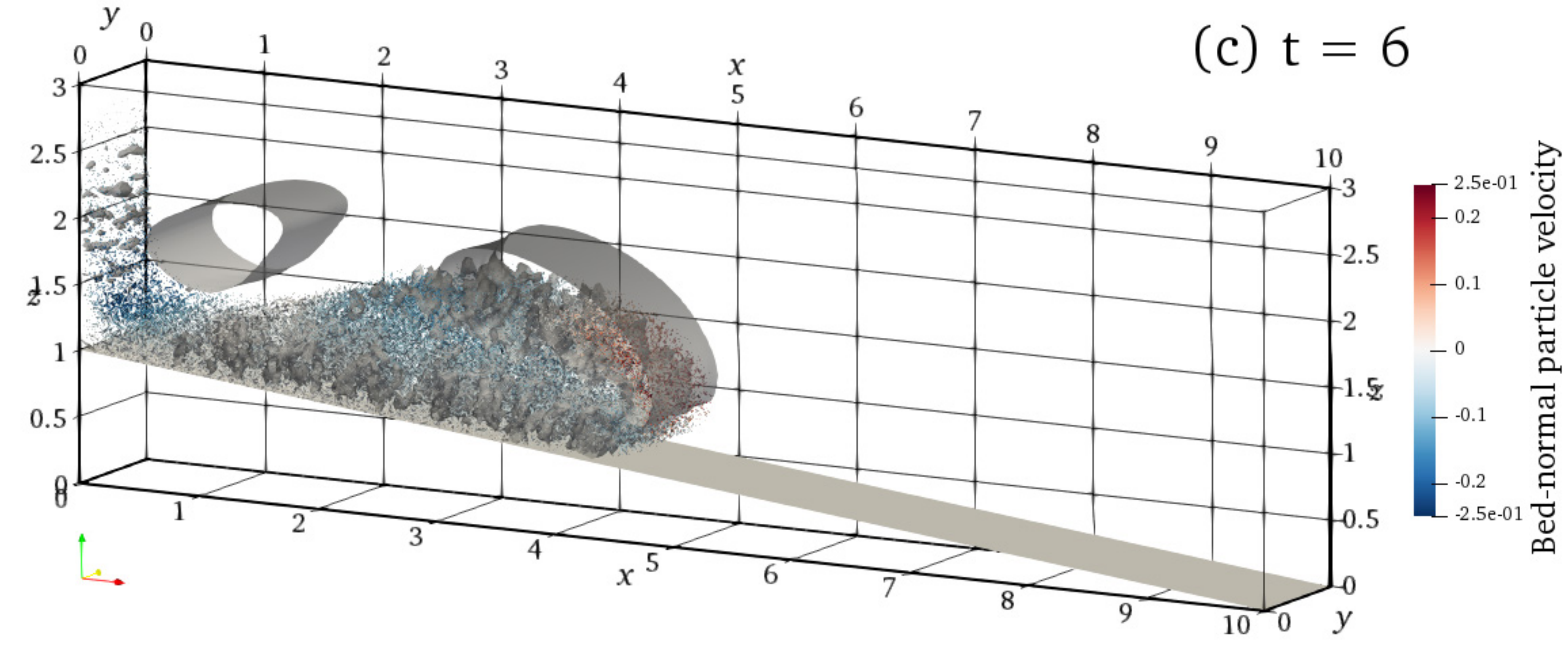}} \\
\centerline{\includegraphics[width=0.7\textwidth]{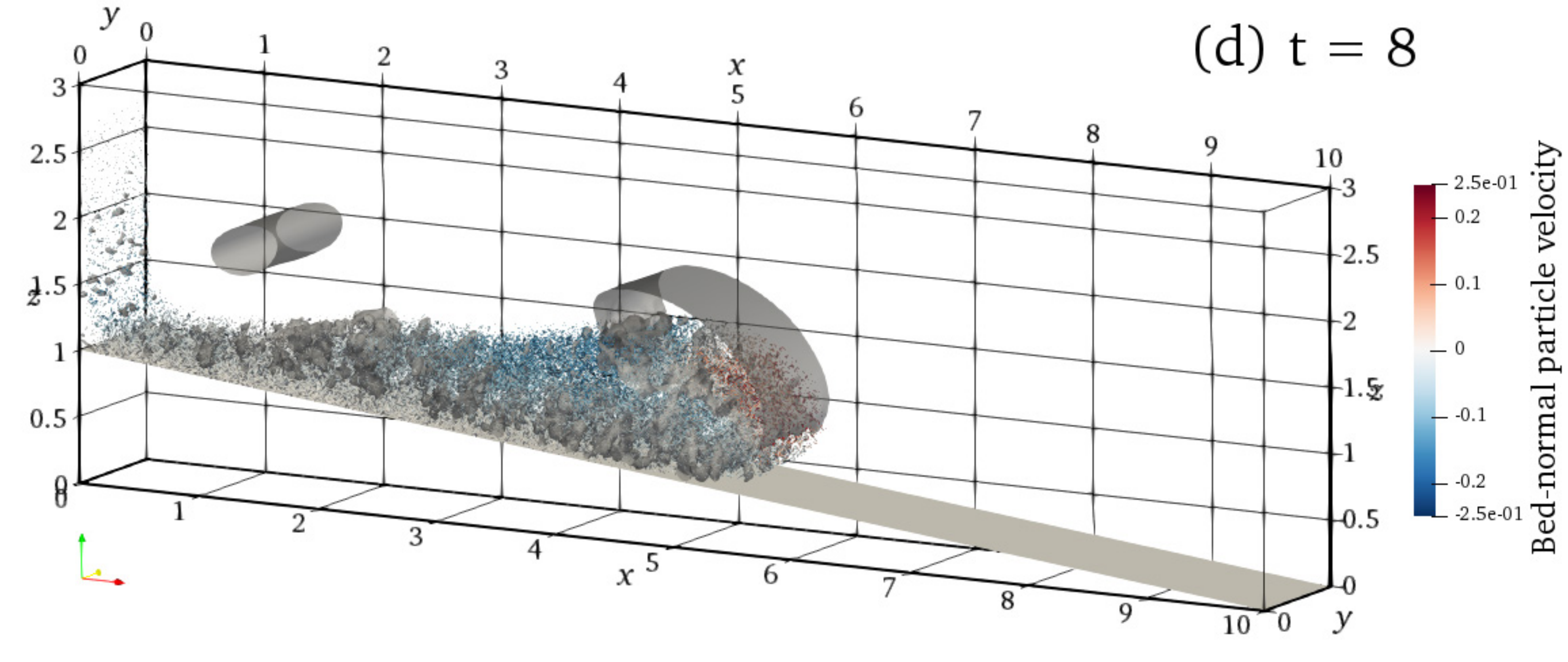}} \\
\centerline{\includegraphics[width=0.7\textwidth]{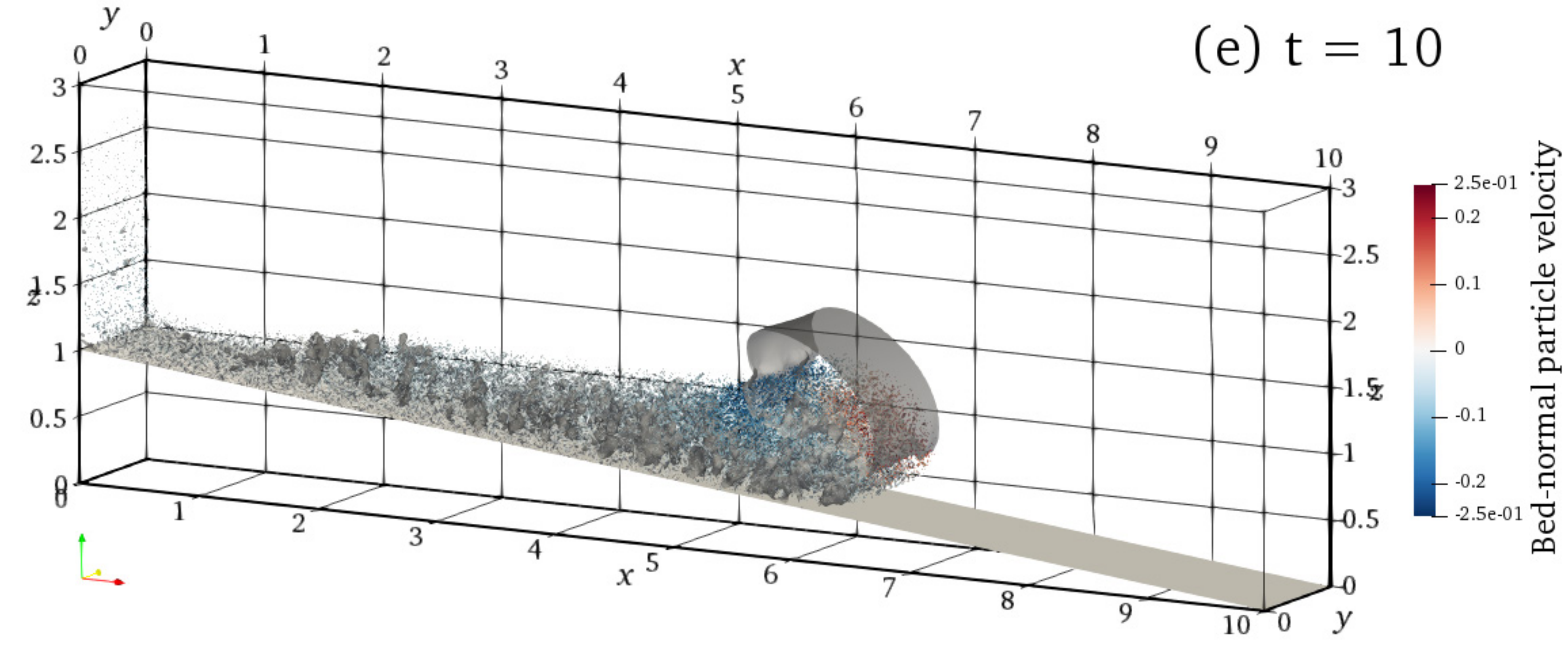}} \\
\end{tabular}
\caption{Coherent vortex structures captured by $Q=0.125$ of case 1 at five selected moments: ($a$) $t=2$, ($b$) $t=4$, ($c$) $t=6$, ($d$) $t=8$ and ($e$) $t=10$. The dispersed particles are also drawn here and rendered with the bed-normal particle velocity.}
\label{fig:Vortical structure for Case 1}
\end{figure}

During the evolution of TCs, typical streamwise and transverse coherent vortical structures are exhibited \citep{dai2022merging}, the qualitative cognition of which is independent of the Reynolds number \citep{nasr2014turbidity, nasr2018mixing}. The coherent structure can also reflect the lobe-and-cleft structures of the TC \citep{espath2015high, francisco2017direct}, which are related to the mixing dynamics. %Since the fluid vortex structures are closely related to the motion of the discrete particles in particle-laden flows \citep{zhu2020interface,crowe1985particle}, 
Here, we attempt to utilize the coherent vortical structures with the $Q$-criterion to survey the suspension regime of the transported particles in TC.
The quantity $Q$ is given by \citep{hunt1988eddies}
\begin{equation}
Q = \frac{1}{2}\left( {{\Omega _{ij}}{\Omega _{ij}} - {S_{ij}}{S_{ij}}} \right),
\label{eq:4_1}
\end{equation}
where the rotation rate tensor $\Omega_{ij}$ and strain rate tensor $S_{ij}$ are respectively expressed as follows:
\begin{equation}
{\Omega _{ij}} = \frac{({u_{f})_{i,j}} - ({u_{f})_{j,i}}}{2},
\label{eq:4_2}
\end{equation}
\begin{equation}
{S_{ij}} = \frac{({u_{f})_{i,j}} + ({u_{f})_{j,i}}}{2}.
\label{eq:4_3}
\end{equation}

Figure \ref{fig:Vortical structure for Case 1} depicts the coherent vortex structures with $Q=0.125$ at five selected moments in case 1, with the dispersed particles coloured with particle bed-normal velocity. Blue particles represent particles that settle toward the wall and red particles represent suspension particles away from the wall.  The settling particles move towards the wall and along the slope due to the advantage of gravity, and the suspension particles are almost in the upper layers of the current head in figure \ref{fig:Vortical structure for Case 1}. 

\begin{figure}
\centerline{\includegraphics[width=\textwidth]{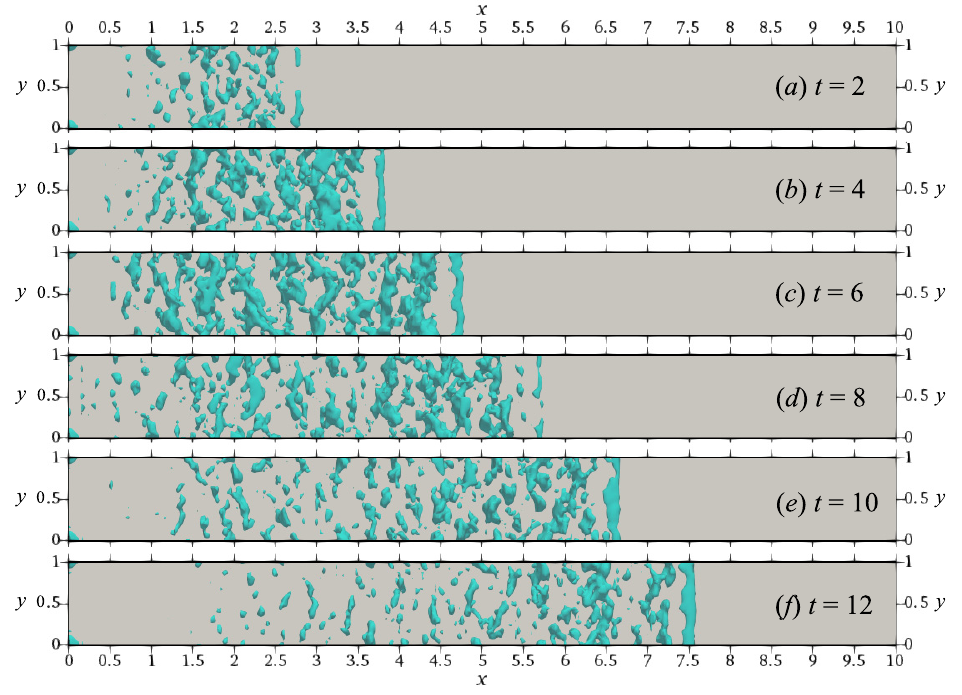}}% Here is how to import EPS art
\caption{\label{fig:figureBottom2} Top-view visualization of the coherent vortex structures below the slice $z'=0.5$ illustrated by the $Q$-criterion with an isovalue $Q=0.25$; ($a$) $t=2$, ($b$) $t=4$, ($c$) $t=6$, ($d$) $t=8$, ($e$) $t=10$ and ($f$) $t=12$.}
\end{figure}

A structurally complete and large-sized structure emerges near the upper interface of the TC in the early stages in figure \ref{fig:Vortical structure for Case 1}($a,b$), then gradually divides into two parts following the time evolution of the current in figure \ref{fig:Vortical structure for Case 1}($c,d$). The front vortex near the head moves forward with the front current during the whole simulation. The vortex at the back stays in place, separates from the mixing layer, shrinks and then vanishes eventually.
The time evolution of the vortex structures is similar to the TC over a flat bed in our previous study \citep{xie2022fluid}. Notably, most of the auto-suspension particles are within the large-scale coherent vortical structure of the head traversing the simulation domain, where $Q>0$ suggests the dominance of the fluid rotation. Thus, it is reasonable to speculate that the uplift of these particles is closely linked to the strong counterclockwise flow around the head, which is consistent with the previous studies \citep{kyrousi2018large,lee2019multi}. The fragmented vortical structures are found in the near-wall region and the mixing layer inside the front vortex. These vortical structures are a manifestation of the combined effects of a complex fluid dynamics with the no-slip condition on the bed \citep{koohandaz2020prediction} and the interaction of particle groups with fluids. In the near-wall region, the height of this vortex structure is generally limited to about 0.5 dimensionless lengths due to the particle transport along the way in the middle layers, and is limited by the upper interface of the TC. 
To facilitate the observation of the coherent structure, top-view coherent vortex structures without particles in the near-wall region below the slice $z'=0.5$ are plotted in figure \ref{fig:figureBottom2}. Relatively large-scale vortical structures can be seen in the near-wall region near the head, which are induced by the intense dynamics of the fluid and particles. During the early stage of the downstream propagation ($t=0\sim6$), small vortex structures of the current (mainly at $x=0\sim2$) gradually increase. The main reason is that the settling particle groups  release their gravitational potential energy to enhance the local fluid kinetic energy, which facilitates the particles advancing.  At $t>6$ (figures \ref{fig:figureBottom2}($c$-$f$)), the small vortex structures in the near-wall region behind the current gradually decline and disappear as a result of dissipation.
In addition, the streamwise development of the coherent vortical structures here is limited due to the small $\Rey$, and the flow is mainly engaged by the transverse vortical structures, agreeing with the understandings of \cite{nasr2014turbidity} and \cite{,koohandaz2020prediction}.

The inflectional instabilities akin to the inviscid Kelvin-Helmholtz mechanism \citep{pelmard2020statistical,ooi20072d} are not observed in figure \ref{fig:Vortical structure for Case 1}. This is because the Richardson number in the present work is $Ri=g'H/U^2\sim O(10)$, where $g'=(\rho_m-\rho_f)|\mathbf{g}|/\rho_f$ is the effective gravitational acceleration with $\rho_m$ denoting the particle--water mixture density, and $U$ the fluid layer-averaged bed-parallel velocity of the TC. The value is much larger than the critical $Ri_c$, which is of the order of unity \citep{turner1986turbulent} even for changes in slope angle and the presence of particles \citep{khavasi2019linear,darabian2021numerical}. Furthermore, due to the low Reynolds number of the TCs in this paper, the viscosity of the fluid is not negligible, which in turn suppresses the occurrence of the inflectional instabilities \citep{khavasi2019linear}.

%A magnified view of the front  for figure \ref{fig:Vortical structure for Case 1} ($b$) is shown in figure \ref{fig:loacl Vortical structure for Case 1}. The two groups of fragment vortical structures mentioned before can be easily distinguished. 

%\begin{figure}
%\centerline{\includegraphics[width=4in]{localt6.pdf}}% Here is how to import EPS art
%\caption{\label{fig:loacl Vortical structure for Case 1} dsdasdas.}
%\end{figure}

%\begin{figure}
%\begin{tabular}{c}

%\end{tabular}
%\end{figure}

\subsection{\label{subsec:Suspension incentives from force analysis}Auto-suspension particle statistics and fluid-particle interactions}

\begin{figure}
\centerline{\includegraphics[width=\textwidth]{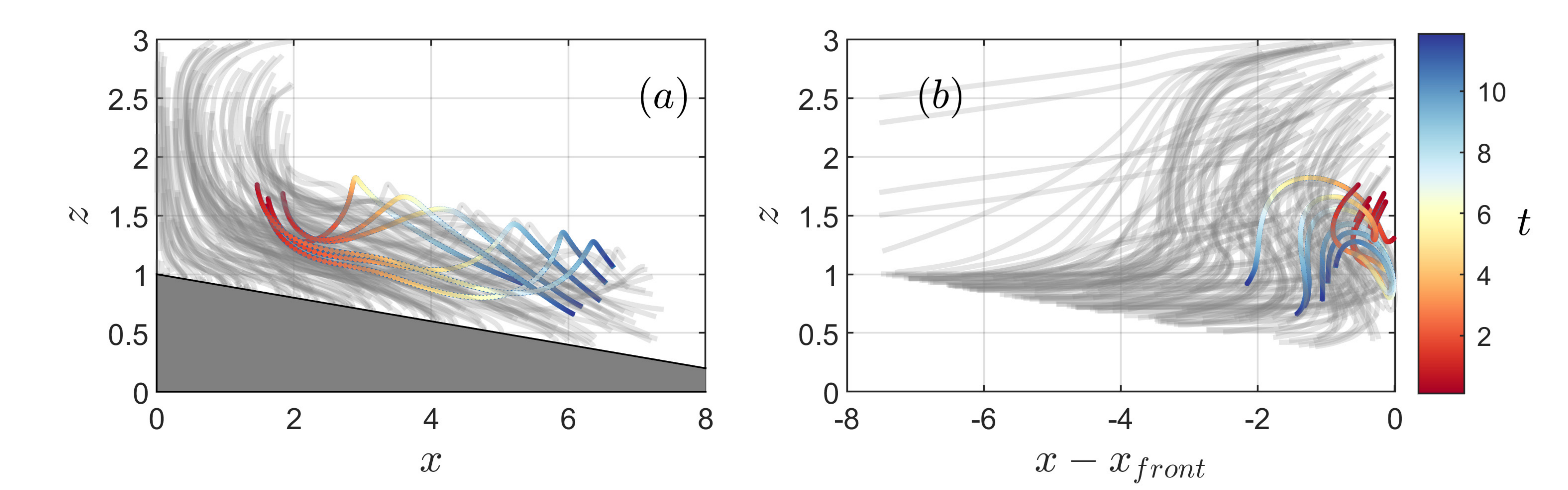}}% Here is how to import EPS art
\caption{\label{fig:Trajectory} Trajectory of 160 particles ($a$) in the tank reference system and ($b$) in the reference frame moving with the TC head. The trajectories of the six representative particles which have entered auto-suspension state are distinguished by coloured lines.}
\end{figure}

%The analysis of particle trajectories helps t
To understand the kinematics of auto-suspension particles in the evolution of TCs, discrete particle motion is discussed in the following section. Figure \ref{fig:Trajectory}($a,b$) directly visualizes the trajectories of 160 particles of case 1 in the tank reference frame and in the reference frame with the moving head, respectively. In these two panels, the six auto-suspension particles selected are highlighted by coloured lines. The whole process of auto-suspension mainly occurs in the head ($x-x_{front}>-2$) as shown in figure \ref{fig:Trajectory}($b$).
The trajectories of the six auto-suspension particles can be described as below:
\begin{enumerate}
\item Particles settle and transport downstream with a constant distance from the slope, as shown in figure \ref{fig:Trajectory}($a$), and they come to go to the front in figure \ref{fig:Trajectory}($b$). 
\item Particles start moving away from the slope (figure \ref{fig:Trajectory}($a$)) and the front (figure \ref{fig:Trajectory}($b$)) during the transport downstream and reach peak points away from the slope. 
\item Particles come to the end of their auto-suspension stage, settle and finally enter the lower layers of the current.
\end{enumerate}

%\begin{figure}
%\centerline{\includegraphics[width=0.6\textwidth]{Figure62.pdf}}% Here is how to import EPS art
%\caption{\label{fig:initialAS} The spatial distribution of the initial positions of the particles that enter auto-suspension, and the particles are rendered with the auto-suspension duration $t_{AS}$.}
%\end{figure}

%Figure \ref{fig:initialAS} shows the spatial distribution of the initial positions of the particles that go into auto-suspension in Case 1, and the auto-suspension duration of each particle. The initial positions of these auto-suspension particles are mainly located in the region of $1.5<x<2$ and $1<z<2$. Particles here are initially affected by the strong annular flow near the interface between the ambient fluid and the fluid-particle mixture, and then evolve to become part of the head of the turbidity current. The particles in the center of this region are capable of maintaining the auto-suspension state for a longer duration.

\begin{figure}
\centerline{\includegraphics[width=\textwidth]{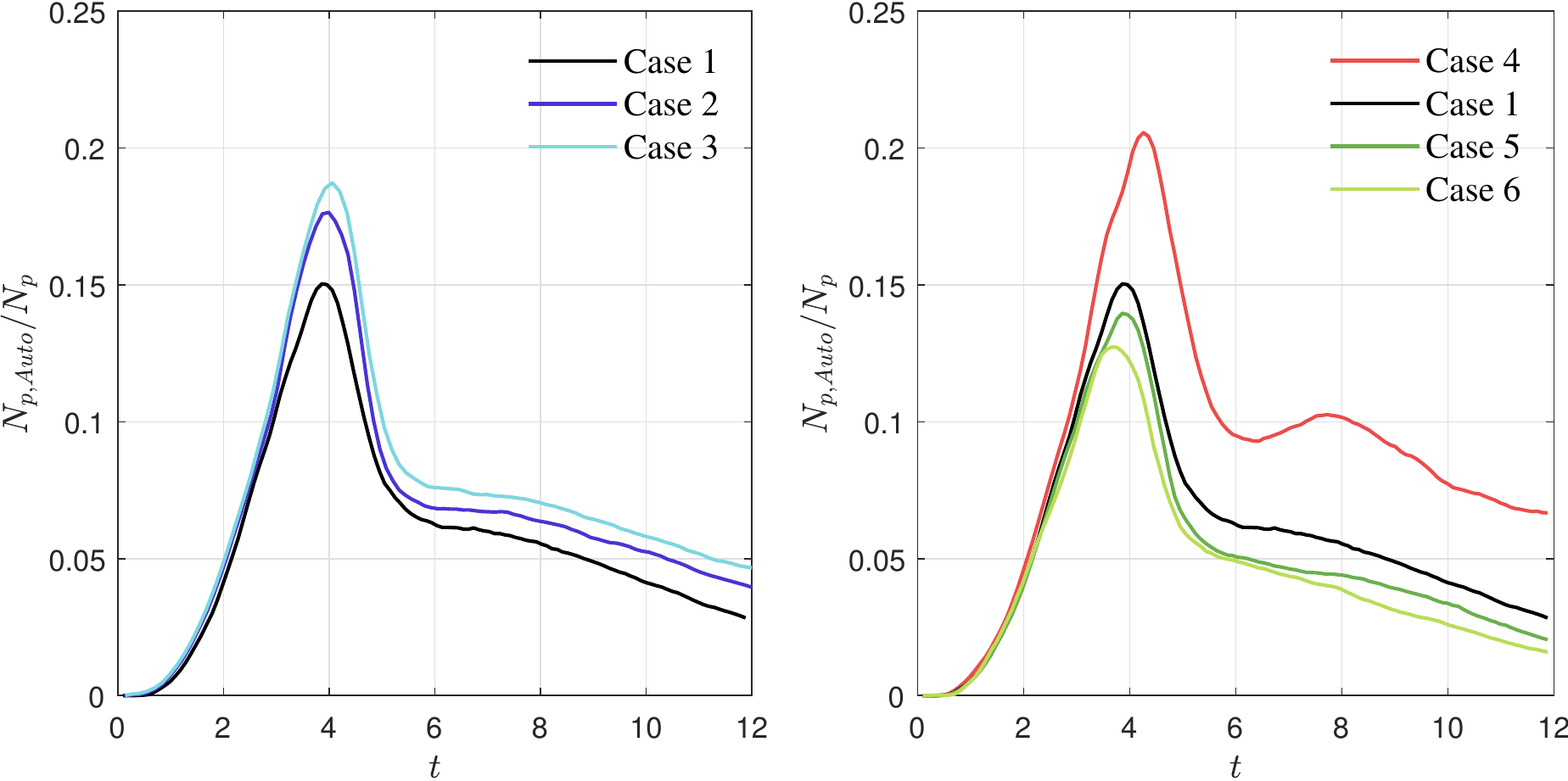}}% Here is how to import EPS art
\caption{\label{fig:ASnumber} Time evolution of the auto-suspension particle quantity. Here, $N_p$ represents the number of all particles, and $N_{p,Auto}$ denotes the number of auto-suspension particles.}
\end{figure}

Figure \ref{fig:ASnumber} plots the time evolution of the auto-suspension particle quantity for all cases. The auto-suspension particle number in all cases increases at $t<4$ and reaches the maximum around $t=4$. Then the value decreases rapidly before $t\approx6$ and diminishes slowly until the end of the simulation, except for case 4 with a relatively large slope angle $\tan\theta=1/5$, where it increases slightly in the range $t=6\sim8$. The possible reason is tightly related to the front vortex, which can carry the particles away from the slope at current head. In case 1, the front vortex shown in figure \ref{fig:Vortical structure for Case 1} is enhanced before $t=4$, then divides into two vortices at $t=6$ and diminishes gradually after that. 

\begin{figure}
\centerline{\includegraphics[width=\textwidth]{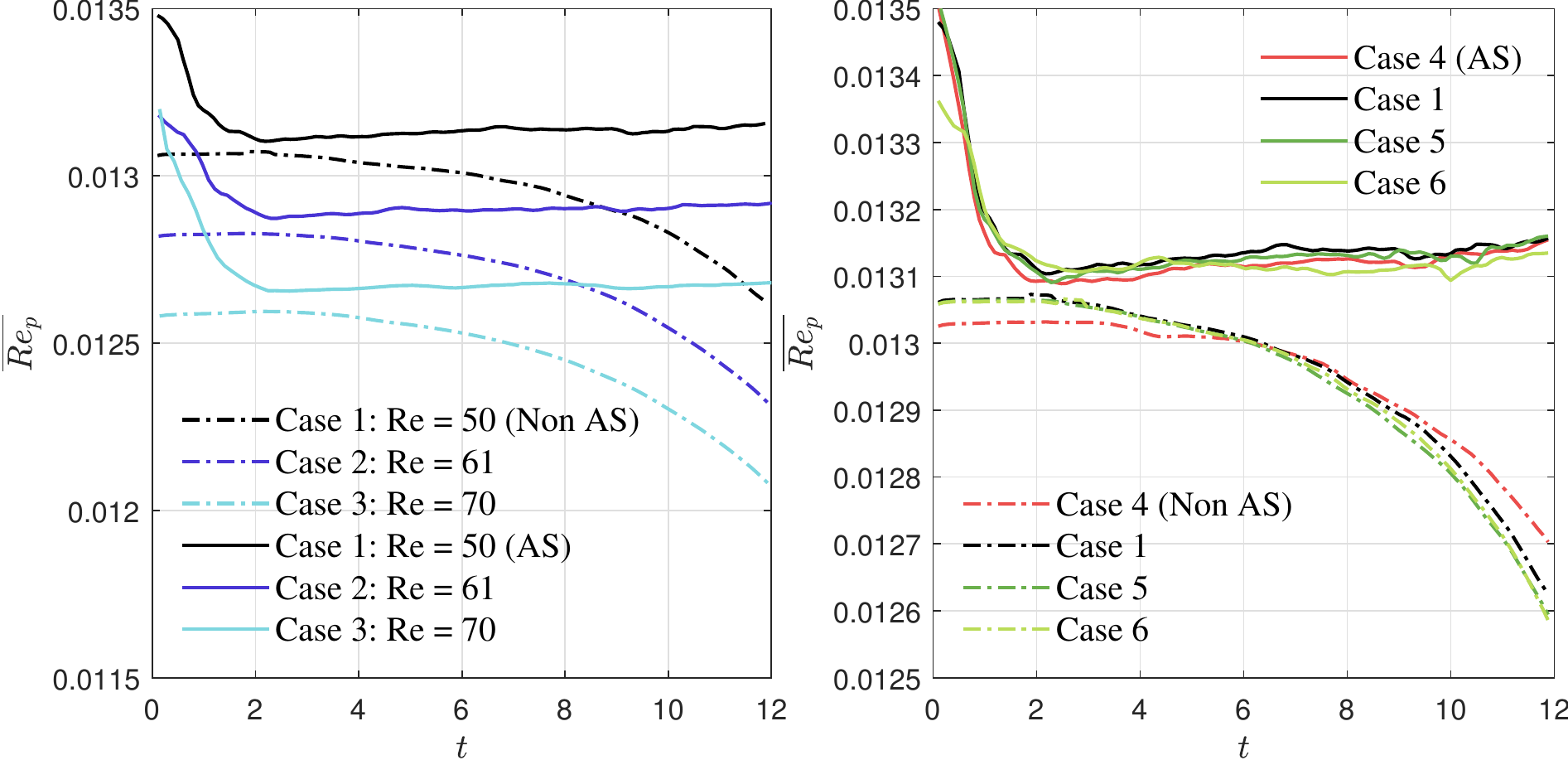}}% Here is how to import EPS art
\caption{\label{fig:Rep} Time evolution of the average particle Reynolds number $\overline{\Rey_p}$ for different cases. The abbreviation “AS” represents “auto-suspension”.}
\end{figure}

The statistics of particle Reynolds numbers $\Rey_p$ are employed to understand the motion patterns of auto-suspension particles in TCs. Substituting the definition of the particle Reynolds number (\ref{eq:11}) into the definition of the particle Stokes number gives $St=\rho_p\Rey_p/(9\rho_f)$, which means that $\Rey_p$ and $St$ are proportional. We compute the average particle Reynolds number $\overline{\Rey_p}$ for the auto-suspension particles and other non-auto-suspension particles in transport, the time evolutions of which are illustrated in figure \ref{fig:Rep}.
It is obviously shown that the average auto-suspension particle Reynolds number is always larger than the non-auto-suspension one, which implies that the auto-suspension particles have a larger $|\mathbf{u}_f-\mathbf{u}_p|$ and suffer a larger drag force.
The average particle Reynolds number of auto-suspension particles decreases during the beginning stage at $t=0\sim2$, and then remains approximately constant (with a slight increase). This indicates that the auto-suspension particles, on average, can maintain a fairly stable motion state during the downstream propagation of the TC. For the non-auto-suspension particles in transport, the $\overline{\Rey_p}$ remains nearly constant at $t=0\sim2$, and subsequently decreases. In figure \ref{fig:Rep}($a$), as particle concentration increases ( $\Rey$ increases from case 1 to case 3), $\overline{\Rey_p}$ becomes smaller, which means the slip velocity between the fluid and particles decreases. This phenomenon is also observed in \cite{sun2022probability}. Figure \ref{fig:Rep}($b$) shows that the differences of $\overline{\Rey_p}$ are quite small, which means the changes in slope (cases 4,1,5,6, with same $\Rey$) do not affect $\overline{\Rey_p}$.

The total force $\mathbf{F}^T$ acting on the particle in the TC comprises the fluid-particle interaction force $\mathbf{F}^f$, the gravity force $\mathbf{G}$ and particle contact force $\mathbf{F}^c$ (the contact force here is the sum of all contact forces experienced by the particle), and is calculated by
\begin{equation}
    \mathbf{F}^T = \mathbf{F}^f+\mathbf{G}+\mathbf{F}^c = \left(
    \mathbf{F}^b + \mathbf{F}^d + \mathbf{F}^l + \mathbf{F}^{add}
    \right)+\mathbf{G} + \mathbf{F}^c.
    \label{eq:24}
\end{equation}

\begin{figure}
\centerline{\includegraphics[width=\textwidth]{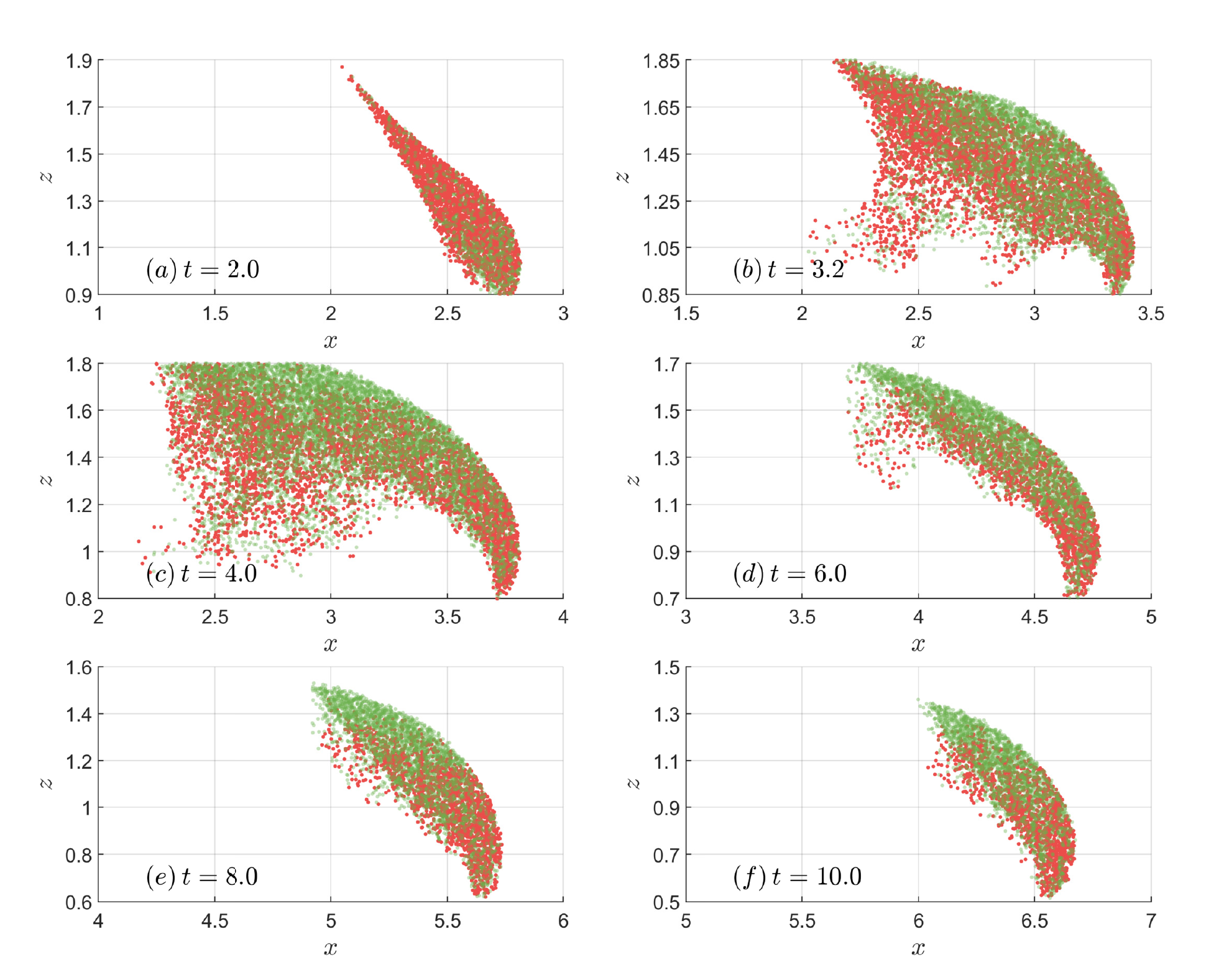}}% Here is how to import EPS art
\caption{\label{fig:Spatial distribution of auto-suspension particles} Spatial distribution of auto-suspension particles at six selected times. Green dot indicates the negative $\mathbf{F}^T_\bot$ particle, while red dot indicates the positive $\mathbf{F}^T_\bot$ particle.}
\end{figure}

Bed-normal total force $\mathbf{F}^T_\bot$ is employed to explain the mechanism of the auto-suspension particles. The spatial distributions of $\mathbf{F}^T_\bot$ of the auto-suspension particles are presented in figure \ref{fig:Spatial distribution of auto-suspension particles}. 
In the early stage at $t=2.0$, the spatial distribution of particles with a positive total force highly coincides with that with a negative total force. As the current evolves, the auto-suspension region expands rapidly and the positive force region separates from the negative one. The particles with positive total force are more widely dispersed in the lower layers, whereas those with negative total force tend to be distributed in the upper layers and closer to the TC profile. It is not difficult to conclude that, when the auto-suspension particles rise and enter the upper zone, they are more likely to reach a turning point and then exit the auto-suspension stage, as shown in figure \ref{fig:Trajectory}. As the annular flow weakens at $t>4.0$, the spatial distribution of particles gradually narrows.

\begin{figure}
\centerline{\includegraphics[width=\textwidth]{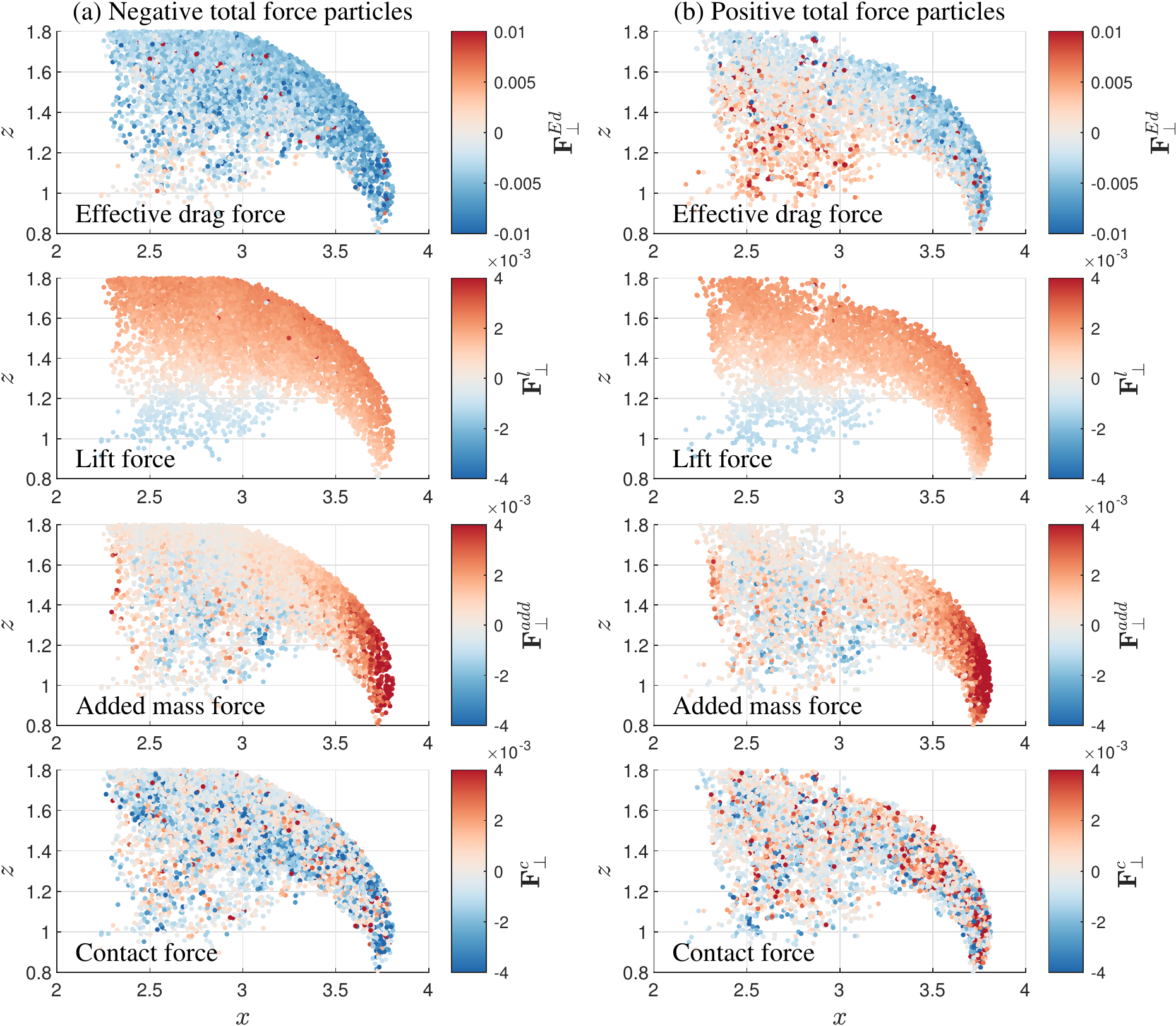}}% Here is how to import EPS art
\caption{\label{fig:Four bed-normal force components} Four dimensionless bed-normal force components ($\mathbf{F}^{Ed}_\bot$, $\mathbf{F}^l_\bot$, $\mathbf{F}^{add}_\bot$ and $\mathbf{F}^c_\bot$) of auto-suspension particles at $t=4.0$. (\textit{a}) The negative $\mathbf{F}^T_\bot$ particles and (\textit{b}) the positive $\mathbf{F}^T_\bot$ particles.}
\end{figure}

Analysing the TC auto-suspension process by describing the spatial variations of each predominant force is both intriguing and practical. The forces in this work are all non-dimensionalized by dividing with $|\mathbf{G}'|$. Figure \ref{fig:Four bed-normal force components}
shows four different dimensionless bed-normal force components of auto-suspension particles at the dimensionless time 4.0.  The effective drag force $\mathbf{F}^{Ed}_\bot$, lift force $\mathbf{F}^l_\bot$, added mass force $\mathbf{F}^{add}_\bot$ and contact force $\mathbf{F}^c_\bot$ are taken into consideration. The bed-normal effective gravity $\mathbf{G}'$ is the main cause of particle settling, and the drag force $\mathbf{F}^d_\bot$ is the most important resistance to settling ($\mathbf{F}^d_\bot/|\mathbf{G}'|\backsim \textit{O}(1)$ and $|\mathbf{G}'\cos\theta|/|\mathbf{G}'|\backsim \textit{O}(1)$). Compared with these two, however, the magnitude of the other forces is relatively small. Thus, we consider the resultant force as the effective drag force $\mathbf{F}^{Ed}_\bot=\mathbf{F}^d_\bot+\mathbf{G}'\cos\theta$ and the magnitude is comparable to the other force components. 

Comparing negative and positive total force particles in figure \ref{fig:Four bed-normal force components}, whether it is the lift force or added mass force, their behaviours on negative and positive $\mathbf{F}^T_\bot$ particles are quite similar. The positive lift force of auto-suspension particles near the TC profile is induced by the positive vorticity near the profile and the positive slip velocity perpendicular to slope. Compared with such a lift force, the added mass force is mostly considerably smaller, except when the particle is very close to the front where the added mass force exhibits an overwhelming positive value. This is due to the strong upward flow at the front. However, the effective drag force and contact force show significant differences. The effective drag forces of negative $\mathbf{F}^T_\bot$ particles are mostly negative, whereas those of positive $\mathbf{F}^T_\bot$ particles are negative near the TC profile area of the head and positive near the central area of the head. The contact force in figure \ref{fig:Four bed-normal force components}(\textit{b}) is almost larger than that in figure \ref{fig:Four bed-normal force components}(\textit{a}) in the perpendicular direction away from the slope. Collision processes can be approximately regarded as the reason for the positive contact force on positive $\mathbf{F}^T_\bot$ particles and the negative contact force on negative $\mathbf{F}^T_\bot$ particles, on average. As a consequence, the bed-normal total force is largely dominated by the effective drag force and contact force. 

Combined with the movement trajectory of particles in figure \ref{fig:Trajectory} and figure \ref{fig:Vortical structure for Case 1}, the fluid vortex at the current head, indicating the upward flow, drives the particles upward away from the slope and the particles enter the auto-suspension state. As the particles continue to rise, the fluid--particle interaction force provided by the flow and the particle--particle contact force cannot resist the particle gravity, leading to a negative $\mathbf{F}^T_\bot$ in figure \ref{fig:Four bed-normal force components}. It prevents the particles from rising further, and the particles eventually reach their peak points. Henceforth, the particles withdraw from this auto-suspension event. Subsequently, particles gradually settle and enter the lower layers of the current.
In essence, the long-term and long-distance cyclic work of the motion system, in which particles continuously enter and exit the auto-suspension state, allows TC to maintain the auto-suspension state from a macro-viewpoint for long distances.

Understanding the average forces on the transported head particles helps us fully explore the auto-suspension mechanism. The transported head particles can be divided into two types: deposited particles and transported particles. The former are very close (within $1.5d_p$) to the bottom wall or deposited particles and their velocity is very small ($|\mathbf{u}_p|<10^{-4}$ m/s), and the latter are the remaining particles. The head average force is obtained by summing the forces acting on transported (the latter type) particles in the head and dividing by the number of corresponding particles. 

\begin{figure}
\centerline{\includegraphics[width=\textwidth]{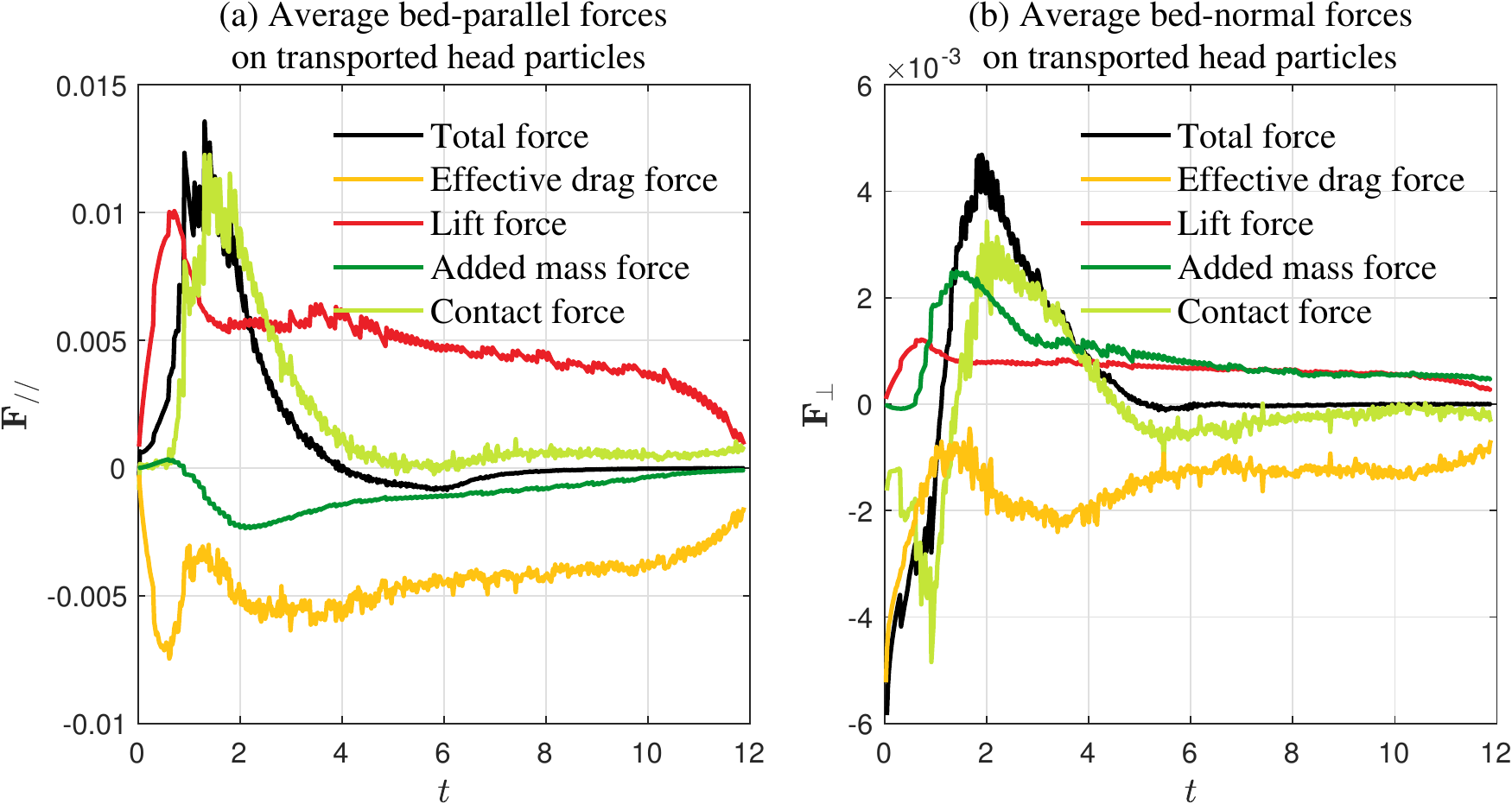}}% Here is how to import EPS art
\caption{\label{fig:forces for case 1} Temporal evolution of (\textit{a}) average bed-parallel forces and (\textit{b}) average bed-normal forces on transported head particles.}
\end{figure}

Figure \ref{fig:forces for case 1}
 shows the temporal evolution of average bed-parallel and bed-normal forces on transported head particles for case 1, including the total force $\mathbf{F}^T$, effective drag force $\mathbf{F}^{Ed}$, lift force $\mathbf{F}^l$, added mass force $\mathbf{F}^{add}$ and contact force $\mathbf{F}^{c}$. In the two directions, given that the drag force components $\mathbf{F}^{d}_{//,\bot}$ and effective gravity components $\mathbf{G}'_{//,\bot}$ are overwhelming and far greater than the other forces, we employ the effective drag force $\mathbf{F}^{Ed}_{//,\bot}$. For the accelerating particles in figure \ref{fig:forces for case 1}(\textit{a}), the positive average $\mathbf{F}^{T}_{//}$ first increases and then decreases at $t=0\sim4$, due to the sum of the lift force and contact force exceeding the sum of the negative effective drag force and added mass force. After $t=4$, $\mathbf{F}^{T}_{//}$ is basically near zero, suggesting that the head particles are roughly advancing at a constant speed on average, consistent with the evolution of the front position \citep{blanchette2005high,he2018investigations}. Due to the fundamental driving action of $\mathbf{G}'_{//}$, the particle velocity is always greater than the fluid velocity, resulting in a negative $\mathbf{F}^{d}_{//}$. The negative $\mathbf{F}^{Ed}_{//}$ in figure \ref{fig:forces for case 1}(\textit{a}) means that the negative drag force is greater than the effective gravity along the slope. %Interestingly, the evolutions of the drag force on a flat slope (Xie et al., 2022) and the effective drag force on an inclined slope here are highly similar.
The average $\mathbf{F}^{l}_{//}$ of the head particles is always positive, which is caused by the positive vorticity around the head (figure \ref{fig:Vortical structure for Case 1}) and the positive slip velocity perpendicular to the slope. Moreover, the head particles are impacted by the collision of particles behind the head, whereby they experience a positive contact force on average. Slightly negative $\mathbf{F}^{add}_{//}$ means the downward particle acceleration is slightly larger than the fluid acceleration along the slope.

In the bed-normal direction in figure \ref{fig:forces for case 1}(\textit{b}), the effective drag force $\mathbf{F}^{Ed}_{\bot}$ is negative when the stationary particles start to settle down at the beginning, which results in the negative total force $\mathbf{F}^{T}_{\bot}$. In the mid-term ($t=1.1\sim5$), the head particles receive a positive total force $\mathbf{F}^{T}_{\bot}$ on average. This positive $\mathbf{F}^{T}_{\bot}$ at $t=1.1$ is generated due to the sum of the lift force and added mass force being larger than the sum of the effective drag force and contact force. Subsequently, the positive $\mathbf{F}^{T}_{\bot}$ experiences an increase and then a decrease to zero, mainly prompted by the change in the positive contact force $\mathbf{F}^{c}_{\bot}$. It is worth mentioning that the positive $\mathbf{F}^{add}_{\bot}$ induced by the upward flow around the head plays a comparable role in the bed-normal direction, as can be seen in figure \ref{fig:Four bed-normal force components} as well. 

\begin{figure}
\centerline{\includegraphics[width=\textwidth]{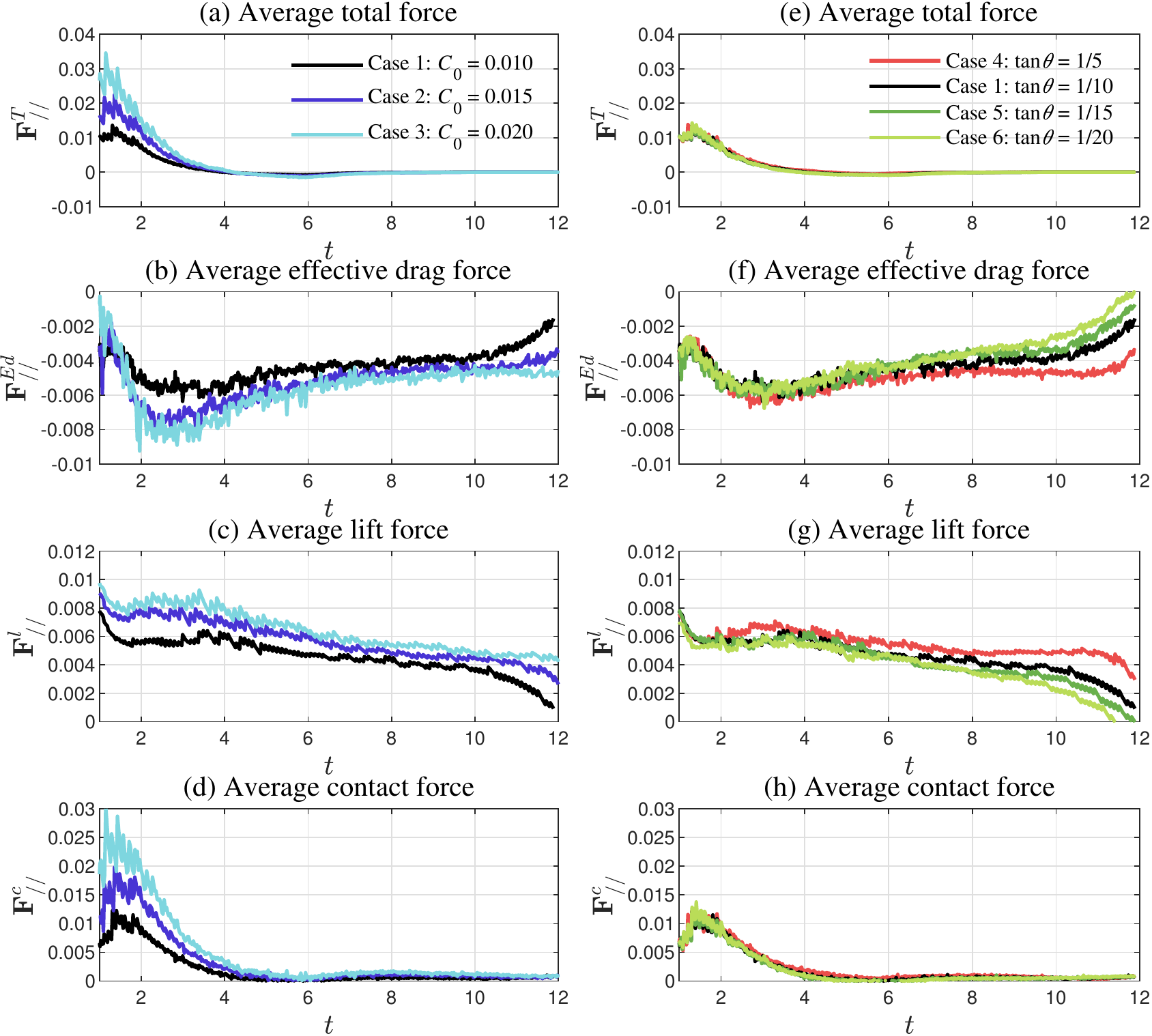}}% Here is how to import EPS art
\caption{\label{fig:15 comparison in parallel} Temporal variation of bed-parallel dimensionless components of (\textit{a} and \textit{e}) average total force, (\textit{b} and \textit{f}) average effective drag force, (\textit{c} and \textit{g}) average lift force and (\textit{d} and \textit{h}) average contact force under different cases.}
\end{figure}

Figure \ref{fig:15 comparison in parallel}
 shows the temporal average variation of the average total force $\mathbf{F}^{T}_{//}$, average effective drag force $\mathbf{F}^{Ed}_{//}$, average lift force $\mathbf{F}^{l}_{//}$ and  average contact force $\mathbf{F}^{c}_{//}$ acting on the transported head particles in our cases. For the same slope, one can easily observe that all the absolute dimensionless force values ($\mathbf{F}^{T}_{//}$, $\mathbf{F}^{Ed}_{//}$, $\mathbf{F}^{l}_{//}$ and $\mathbf{F}^{c}_{//}$) increase with increasing particle concentration in figure \ref{fig:15 comparison in parallel}(\textit{a}-\textit{d}). The increasing $\mathbf{F}^{T}_{//}$ explains that the higher the concentration, the faster the head of the current \citep{hu2020layer}. In figure \ref{fig:15 comparison in parallel}(\textit{e}-\textit{h}), increasing the slope has a low impact on these forces parallel to the slope. Only in the later stage of the duration do the negative effective drag force and the positive lift force increase.

For the three increasing force components in all cases, reasons can be explained as follows:
\begin{enumerate}
\item The absolute $\mathbf{F}^{Ed}_{//}$ increases due to the increasing particle velocity around the head, which is shown in figure \ref{fig:head position}. 
\item The positive $\mathbf{F}^{l}_{//}$ increases due to the increase in positive vorticity around the head. 
\item The positive $\mathbf{F}^{c}_{//}$ increases due to the increasing collision probability of the particles in the current with higher particle concentration. 
\end{enumerate}

\begin{figure}
\centerline{\includegraphics[width=\textwidth]{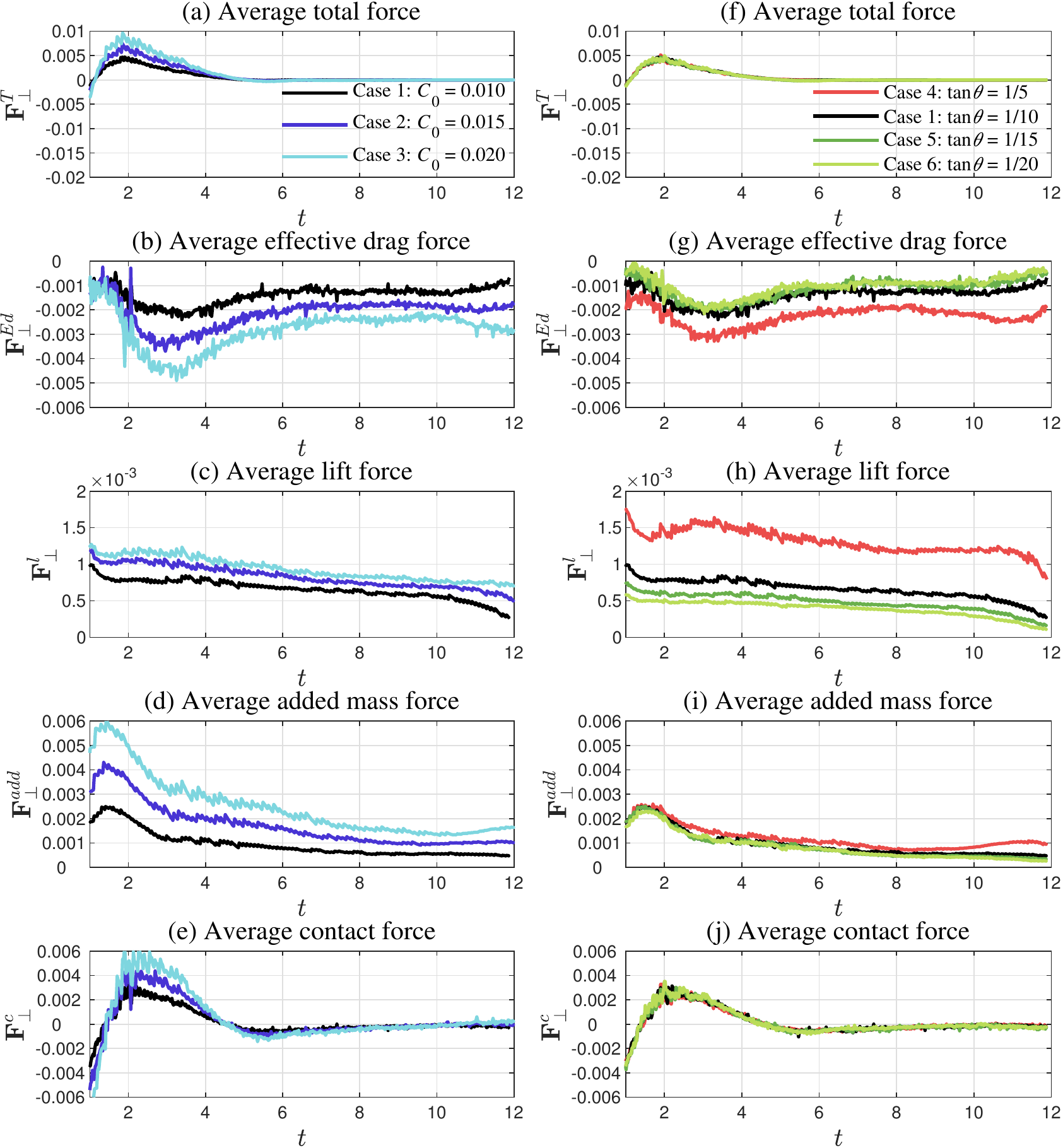}}% Here is how to import EPS art
\caption{\label{fig:16 comparison in perpendicular} Temporal variation of bed-normal dimensionless components of (\textit{a} and \textit{f}) average total force, (\textit{b} and \textit{g}) average effective drag force, (\textit{c} and \textit{h}) average lift force, (\textit{d} and \textit{i}) average added mass force and (\textit{e} and \textit{j}) average contact force under different cases.}
\end{figure}

Figure \ref{fig:16 comparison in perpendicular}
 plots average total force $\mathbf{F}^{T}_{\bot}$, average effective drag force $\mathbf{F}^{Ed}_{\bot}$, average lift force $\mathbf{F}^{l}_{\bot}$, average added mass force $\mathbf{F}^{add}_{\bot}$ and average contact force $\mathbf{F}^{c}_{\bot}$ on transported head particles over time for the different cases. In overall terms, an increase in particle concentration allows an increase in the absolute value of each force perpendicular to the slope, where an increase in the positive total force $\mathbf{F}^{T}_{\bot}$ implies an enhancement in the auto-suspension capacity. This leads to an increase in the collision probability of the particles, which in turn drives a corresponding growth in the contact force $\mathbf{F}^{c}_{\bot}$. Here, $\mathbf{F}^{c}_{\bot}$ and $\mathbf{F}^{add}_{\bot}$ are increased due to the enhanced flow strength. The increase of negative $\mathbf{F}^{Ed}_{\bot}$ reflects a decrease of the positive drag force $\mathbf{F}^{d}_{\bot}$, which is of interest. The greater $\mathbf{F}^{l}_{\bot}$ (figure \ref{fig:16 comparison in perpendicular}(\textit{c})), $\mathbf{F}^{add}_{\bot}$ (figure \ref{fig:16 comparison in perpendicular}(\textit{d})) and $\mathbf{F}^{c}_{\bot}$ (figure \ref{fig:16 comparison in perpendicular}(\textit{e})) provide a stronger effect for particles to suspend and rise up, prompting an increase in positive $u_p^\bot$, which thus results in a decrease in $\mathbf{F}^{d}_{\bot}$. The change in slope has a negligible effect on the total force $\mathbf{F}^{T}_{\bot}$ (figure \ref{fig:16 comparison in perpendicular}(\textit{f})), added mass force $\mathbf{F}^{add}_{\bot}$ (figure \ref{fig:16 comparison in perpendicular}(\textit{i})) and contact force $\mathbf{F}^{c}_{\bot}$ (figure \ref{fig:16 comparison in perpendicular}(\textit{j})). However, it allows the negative effective drag force and the positive lift force to be enhanced (figure \ref{fig:16 comparison in perpendicular}($g, h$)), and the magnitudes of these two changes appear to balance out.

\subsection{\label{subsec:Energy budget}Energy budget}

The energy of the TC includes the particle gravitational potential energy $E_p^p$, the particle kinetic energy $E_k^p$, the fluid potential energy $E_p^f$, the fluid kinetic energy $E_k^f$ and the dissipated energy $E_{Diss}$ \citep{xie2022fluid,zhu2022grain}. We herein concentrate on how the changes in potential and kinetic energies ($\Delta E_p^p$, $\Delta E_k^p$, $\Delta E_p^f$ and $\Delta E_k^f$) evolve throughout the history of the TC, where $\Delta$ represents the change between the energy at the current moment and the energy at the initial moment (e.g., $\Delta E_k^p\left(t\right) = E_k^p\left(t\right) - E_k^p\left(0\right)$, where $0$ is the initial moment). The four kinetic and potential energy components can be calculated as follows:
%Due to energy conservation, the above energy components have the following relationship \citep{xie2022fluid}:
%\begin{equation}
%    \Delta E_p^p\left(t\right) + 
%    \Delta E_k^p\left(t\right) +
%    \Delta E_p^f\left(t\right) + 
%    \Delta E_k^f\left(t\right) + 
%    \Delta E_{Diss}\left(t\right)
%    = 0,
%    \label{eq:28}
%\end{equation}

\begin{equation}
    E_p^p\left(t\right) = 
    \sum\limits_{i=1}^{N_p} m_i \left|\mathbf{g}\right| z_{p,i},
    \label{eq:29energy}
\end{equation}
\begin{equation}
    E_k^p\left(t\right) = 
    \sum\limits_{i=1}^{N_p} \left(
    \frac{1}{2} m_i \left|\mathbf{u}_{p,i}\right|^2 + 
    \frac{1}{2} I \left|\mathbf{\omega}_{p,i}\right|^2
    \right),
    \label{eq:30energy}
\end{equation}
\begin{equation}
    E_p^f\left(t\right) = 
    \int _\Omega \alpha_f \rho_f \left|\mathbf{g}\right| z d V,
    \label{eq:31energy}
\end{equation}
\begin{equation}
    E_k^f\left(t\right) = 
    \int _\Omega \frac{1}{2} \alpha_f \rho_f 
    \left|\mathbf{u}_f\right|^2 d V,
    \label{eq:32energy}
\end{equation}
where $z_{p,i}$ is the elevation of particle $i$ and $\Omega$ represents the whole simulation domain. 
%Thus, the dissipated energy $E_{Diss}$ can be obtained by combining Eqs. (\ref{eq:28})$\sim$(\ref{eq:32energy}). 

\begin{figure}
\centerline{\includegraphics[width=\textwidth]{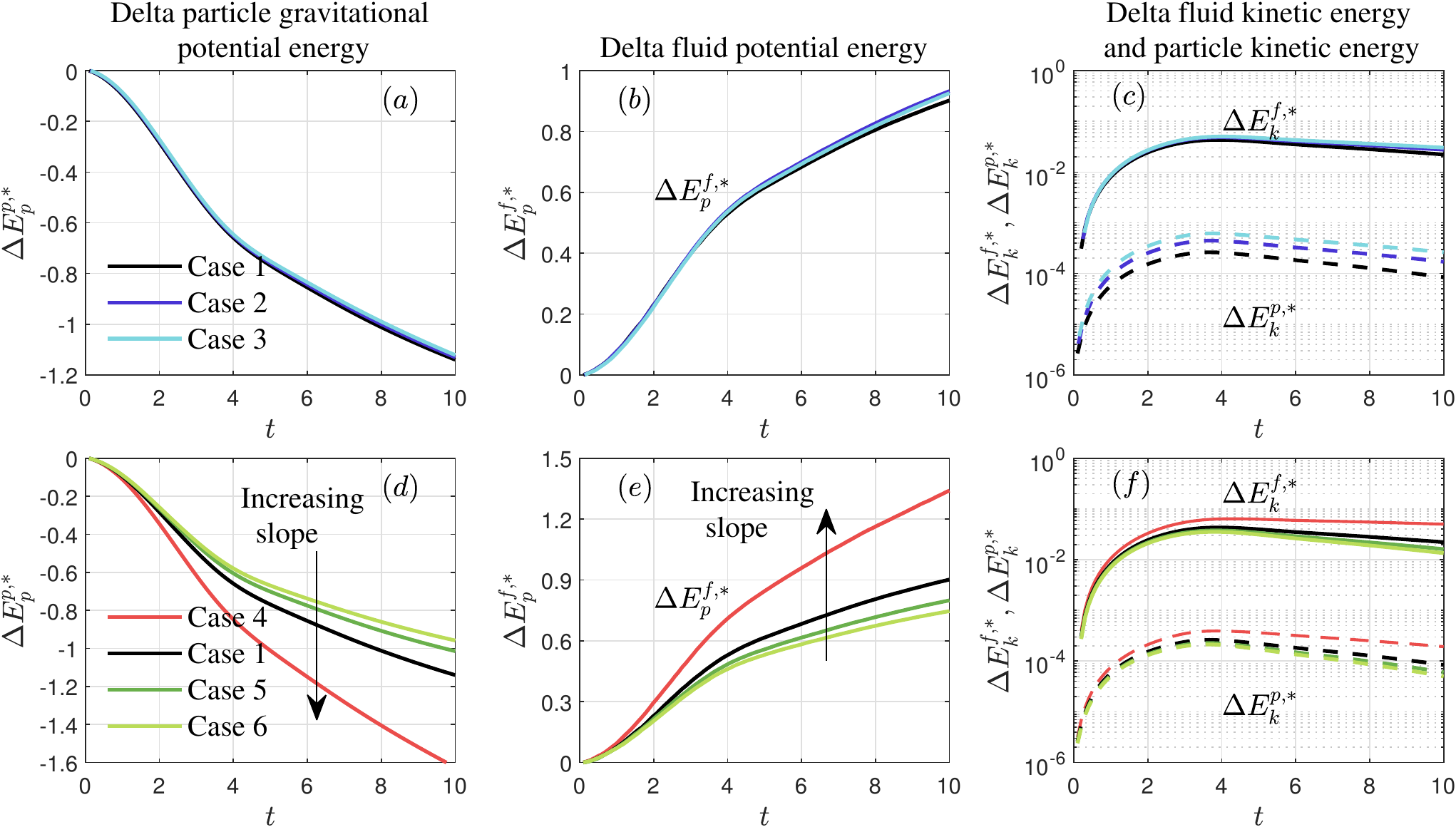}}% Here is how to import EPS art
\caption{\label{fig:17 Energy} Temporal evolution of non-dimensionalized energy components: ($a$ and $d$) $\Delta E_p^{p,\ast}$, ($b$ and $e$) $\Delta E_p^{f,\ast}$, and ($c$ and $f$) $\Delta E_k^{f,\ast}$, $\Delta E_k^{p,\ast}$.}
\end{figure}

Figure \ref{fig:17 Energy} plots the temporal variations of the energy components, non-dimensionalized by the relative initial particle gravitational potential energy depending on the reference plane ($z=L_B$), which is defined as $E_p^p\left(0\right) = \sum\limits_{i=1}^{N_p} m_i \left|\mathbf{g}\right| (z_{p,i}-L_B)$. All energy components with superscript “$\ast$” in the figure are non-dimensionalized. From the perspective of the energy budget, the energy of the TC on the inclined slope is converted from the particle gravitational potential energy into fluid potential energy, fluid kinetic energy, particle kinetic energy and the dissipated energy due to viscosity \citep{dai2015high,he2018investigations}. Defining the available potential energy as $E_{p,avail}=-\Delta E_p^p-\Delta E_p^f$, then at $t=10$, approximately $10\sim20\%$ is converted into fluid and particle kinetic energy, and the rest (approximately $80\sim90\%$) is dissipated due to the fluid viscosity and the particle–particle collisions. Figure \ref{fig:17 Energy}($a$-$c$) shows that as the particle concentration increases, the non-dimensionalized energy components are similar. The increase in particle concentration boosts the conversion of particle kinetic energy, which means a faster particle velocity. This reflects the fact that, the larger the initial particle concentration is, the faster the TC advances \citep{hu2020layer,farizan2019effect}. The possible reason is that, with the increasing particle concentration, all the absolute values of the force components increase (shown in figures \ref{fig:15 comparison in parallel} and \ref{fig:16 comparison in perpendicular}) due to higher collision frequency. 

For different slope angles in cases 1,4,5 and 6, the reference plane is $z=L_B$, which means that the initial particle gravitational potential energies $E_p^p(0)$ are nearly equal.
As shown in figure \ref{fig:17 Energy}($d$-$f$), the increase in slope leads to a significant consumption of the particle gravitational potential energy \citep{steenhauer2017dynamics}, which leads to a considerable increase in the fluid potential energy, fluid kinetic energy and particle kinetic energy \citep{he2018investigations,francisco2017direct}. 

%\section{\label{sec:Effects of two key factors on particle auto-suspension}Effects of two key factors on particle auto-suspension}

%This section discusses the effects of two key factors on auto-suspension, which are initial particle concentration $C_0$ and slope angle $\theta$. Auto-suspension criterion is developed in Subsection \ref{subsec:Criterion for auto-suspension} and bed shear stress is explored in Subsection \ref{subsec:Bed shear stress}. Given that auto-suspension mostly occurs in the head, Subsection \ref{subsec:Head average force} describes the time evolution of the average forces on transported head particles and their reactions to the two key factors. In Subsection \ref{subsec:Energy budget}, the energy budget is explored in different cases.

\subsection{\label{subsec:Criterion for auto-suspension}Criterion for auto-suspension}

The auto-suspension criterion is generally used for predicting the tendency of auto-suspension in TC on a slope. Based on the balance of force and energy, \cite{bagnold1962auto}, \cite{pantin1979interaction} and \cite{parker1982conditions} proposed different criteria for the auto-suspension index of the TC head as follows:
\begin{equation}
    k_{Auto,h}=-\frac{w_{p,h}\cos\theta}{u_{p,h}^{//}\sin\theta}\leq k_T,
    k_T=\left\{
    \begin{array}{cc}
    \cos\theta & \mathrm{(Bagnold, 1962)}\\
    0.01 & \mathrm{(Pantin, 1979)}\\
    1 & \mathrm{(Parker, 1982)}
\end{array}\right.
    \label{eq:25}
\end{equation}
where $k_{Auto,h}$ is the head auto-suspension index, $w_{p,h}$ is the average vertical velocity of head particles, $u_{p,h}^{//}$is the average velocity along the slope for the head particles and $k_T$ is the threshold for auto-suspension criterion. 

\begin{figure}
\centerline{\includegraphics[width=0.85\textwidth]{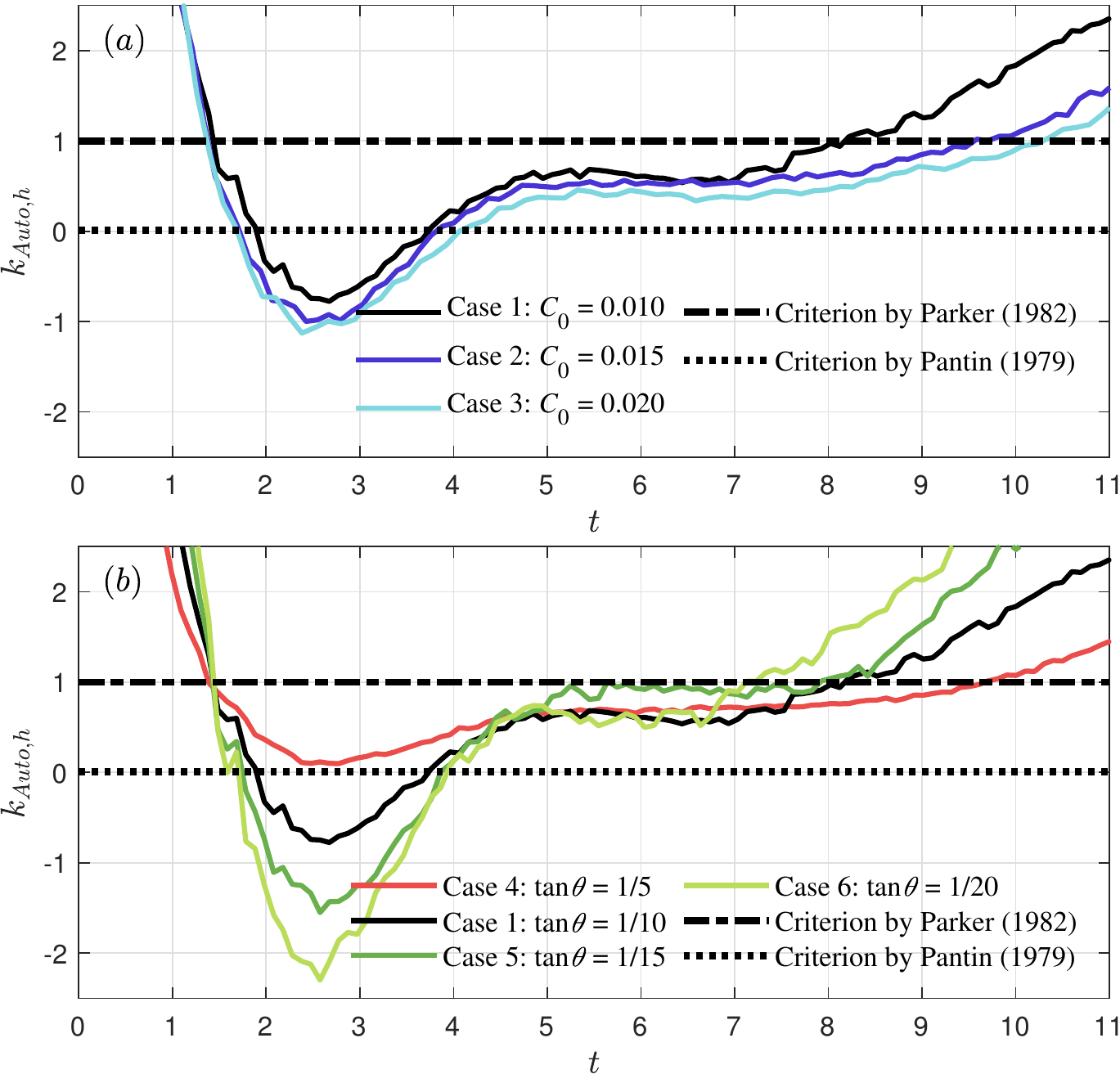}}% Here is how to import EPS art
\caption{\label{fig:auto-suspension index} Temporal evolutions of head auto-suspension index for (\textit{a}) different initial particle concentrations (cases 1, 2 and 3) and for (\textit{b}) different slope angles (cases 4, 1, 5 and 6).}
\end{figure}

The temporal evolutions of the head auto-suspension index under different initial particle concentrations are plotted in figure \ref{fig:auto-suspension index}(\textit{a}), and those for different bottom slopes are plotted in figure \ref{fig:auto-suspension index}(\textit{b}). The criteria proposed by \cite{bagnold1962auto} and \cite{parker1982conditions} differ little for the cases in this work (for case 4, $\cos\theta=0.98$), and thus the lines in figure \ref{fig:auto-suspension index} indicate criteria by \cite{parker1982conditions} and \cite{pantin1979interaction}. Auto-suspension regions are located below the lines while non-auto-suspension regions are above the lines. The auto-suspension index is negative (smaller than the \cite{pantin1979interaction} criterion) only at about $1.8\sim4$ dimensionless time except for case 4 in figure \ref{fig:auto-suspension index}. Actually, the negative auto-suspension index lasts only for a short period of time and auto-suspension particles do exist in case 4. The comparison between the present data and the criteria suggests that the Pantin’s standard might be excessively strict for auto-suspension \citep{sequeiros2009experimental}. Thus we use the \cite{parker1982conditions} criterion hereafter.

It can be seen in figure \ref{fig:auto-suspension index} that the exact moment when the current starts to fulfil Parker's auto-suspension criterion is roughly the same (about $t=1.4$). The time head particles exiting the auto-suspension will be delayed with increasing particle concentration $C_0$ or increasing slope angle $\theta$. This demonstrates that both can extend the period of auto-suspension, exhibiting the behaviour of enhancing the auto-suspension maintenance ability by increasing $C_0$ or $\theta$. 

\begin{figure}
\centerline{\includegraphics[width=\textwidth]{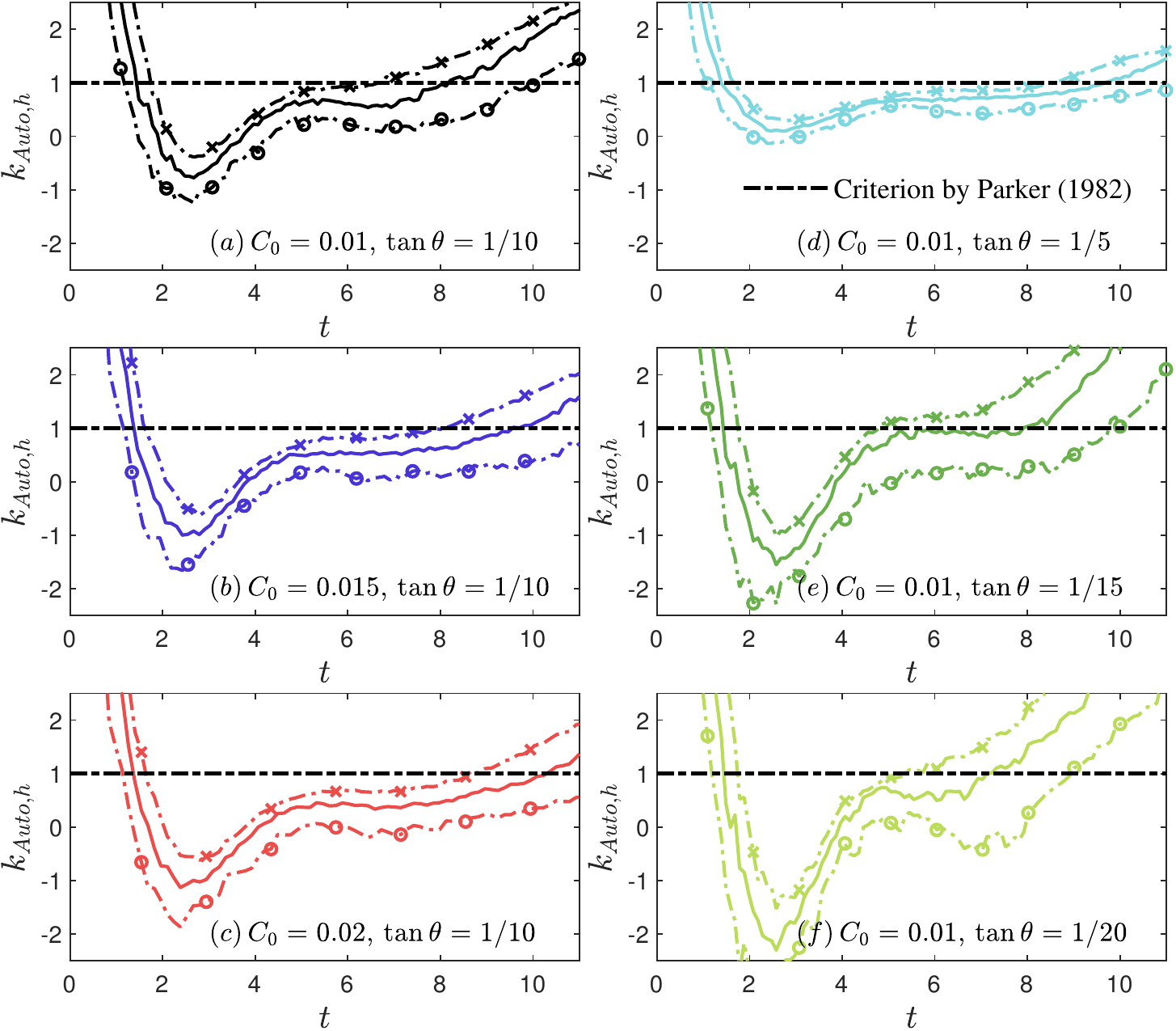}}% Here is how to import EPS art
\caption{\label{fig:auto-suspension sensitivity}Temporal evolutions of head auto-suspension index for various cases with three head length definition methods. The dotted lines with circles represent 0.10 times the total length of the current, the solid lines 0.15 times the total length and the dotted lines with crosses 0.20 times the total length.}
\end{figure}

The impact of different head length definitions on the head auto-suspension index are discussed. Figure \ref{fig:auto-suspension sensitivity} depicts the time evolution of the head auto-suspension index for six cases under three head length definitions (0.10, 0.15 and 0.20 times the total length of the current). As can be observed, different head length definitions in the range of $0.10\sim0.20$ times the TC length do not qualitatively affect the evolution of the head auto-suspension index. However, the head auto-suspension index $k_{Auto,h}$ gets smaller as the head length decreases, extending the duration during which the auto-suspension standard is satisfied. This is because particle auto-suspension mostly occurs near the front.

The depth-averaged auto-suspension index along the slope $k_{Auto,d}$, which also can be said to be the local auto-suspension index, can be expressed, with reference to (\ref{eq:25}), as follows:
\begin{equation}
    k_{Auto,d}=-\frac{w_{p,d}\cos\theta}{u_{p,d}^{//}\sin\theta},
    \label{eq:26}
\end{equation}
where $w_{p,d}$ is the depth-averaged particle vertical velocity and $u_{p,d}^{//}$ denotes the depth-averaged bed-parallel particle velocity. 

\begin{figure}
\centerline{\includegraphics[width=\textwidth]{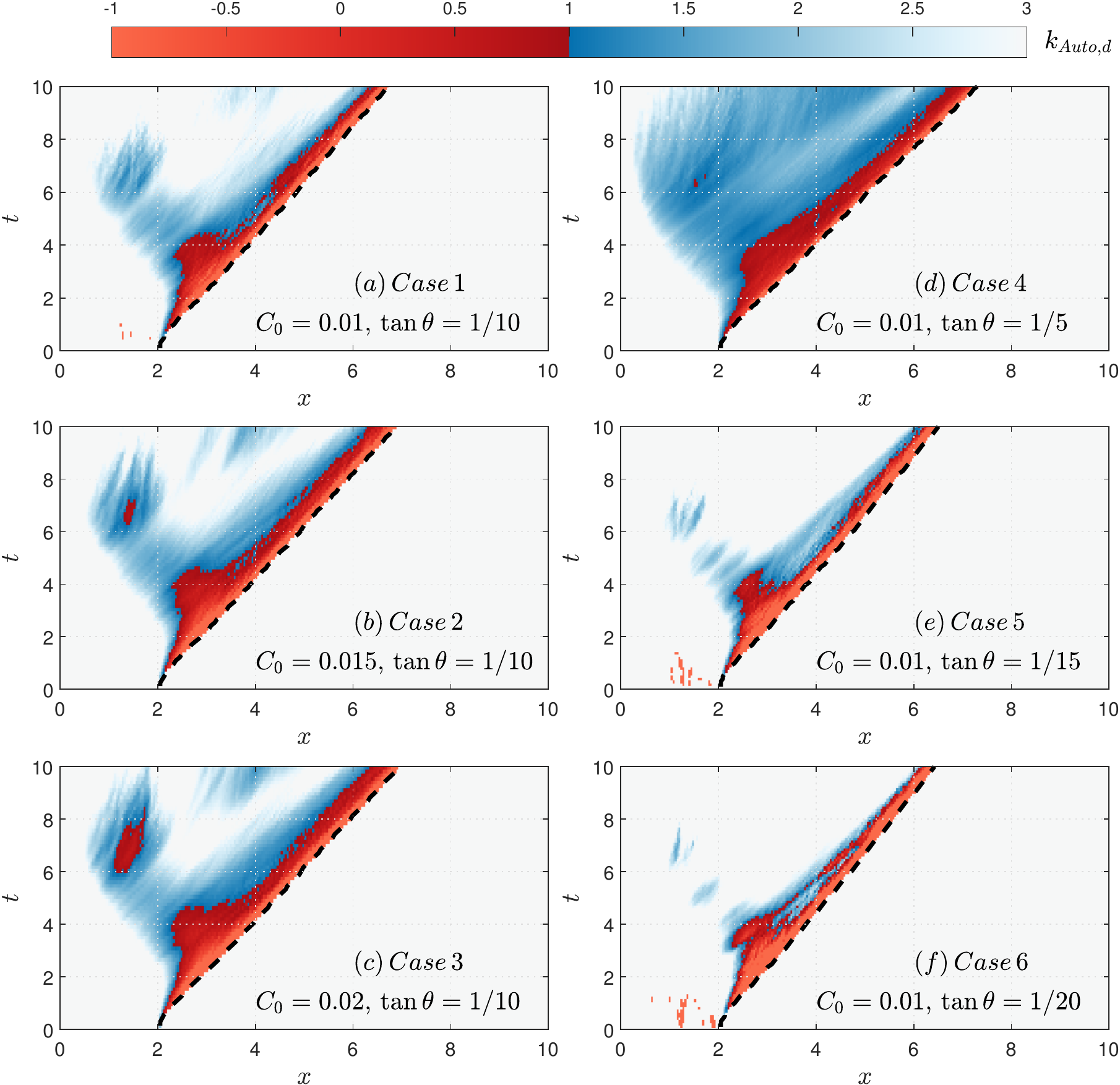}}% Here is how to import EPS art
\caption{\label{fig:auto-suspension index t_x} Temporal and spatial variations of local auto-suspension index for various cases. Black dotted line indicates the front position of the TC.}
\end{figure}

Figure \ref{fig:auto-suspension index t_x}
 shows the temporal and spatial distributions of the local auto-suspension index, with the dashed line representing the TC front. The auto-suspension index, which is distinguished by colour in the figure, is judged by using Parker’s standard (red means particles in auto-suspension and blue means particles tending to auto-suspension). It can be seen that the red areas meeting Parker’s criterion are mainly near the front of the TC, which demonstrates that it is reasonable to focus on the head dynamics when studying the auto-suspension of the TC \citep{sequeiros2009experimental}. The red auto-suspension areas are not sensitive to the particle concentration and slope angle. The blue areas expand 
to some extent when the initial particle concentration or slope rises, especially that area in case 4 shown in figure \ref{fig:auto-suspension index t_x}(\textit{d}) that expands a lot. At about $t=4$, the spatial expansion of the auto-suspension area reaches its maximum, which shows a good agreement with figure \ref{fig:Spatial distribution of auto-suspension particles}. Interestingly, there exists a special blue region away from the dashed line in the range of $x=0\sim2$ at $t=6\sim8$, in which the particles tend to auto-suspension. The main reason might be the influence of the reflected wave generated by the backward flow \citep{nasr2014turbidity,bonnecaze1993particle,blanchette2005high}.

\section{\label{sec:Summary and conclusions}Summary and conclusions}

A TC can transport for long distances on very gentle slopes, which is inseparable from the auto-suspension mechanism behind it. %In the past, researchers mainly analyzed the auto-suspension mechanism of TC from the perspectives of hydrodynamic process, particle transport flux, formed deposit structure, and energy conversion \citep{pantin2011improved,sequeiros2009experimental,sequeiros2018internal}. 
This paper employs the LES-DEM model to simulate lock-exchange TCs on inclined slopes. The feasibility of the LES-DEM model for simulating TCs on inclined slopes is examined though the quantitative comparison of the temporal variation of the front position with the experimental result \citep{gladstone1998experiments}, and the quantitative comparison of the fluid velocity profile with an empirical formula. The auto-suspension mechanism of the current is explained from the perspective of the particle flow process and fluid-particle interactions, and the dynamic response of the auto-suspension process with respect to the two key parameters, initial particle concentration $C_0$ and terrain slope angle $\theta$, is discussed.

%The LES-DEM model is capable of accurately describing the kinematic and dynamic characteristics of each particle. 
The value of $\overline{\Rey_p}$ of all transported particles is negatively correlated with the $\Rey$ of the current. The auto-suspension particles mainly appear near the current head during the current evolution. The auto-suspension particle quantity and area exhibit a tendency to increase first and then decrease, and have a high positive correlation with the coherent vortical structure near the head. The average particle Reynolds number $\overline{\Rey_p}$ of auto-suspension particles can remain approximately unchanged as the TC evolves downstream, and is larger than the non-auto-suspension particle Reynolds number that gradually decreases.

The movements of auto-suspension particles during transport are induced by the complicated physical regimes. The particle trajectories and forces are analysed. The rising auto-suspension particles mainly occur near the current front, reach the peak away from the slope, finally enter the lower layers of the current and exit the auto-suspension state.
For the auto-suspension particles, the bed-normal lift force and added mass force are mostly positive. 
The effective drag force (the combination of effective gravity and drag force) and contact force are much larger in positive total force particles than those in negative ones. Throughout the process, particles are constantly entering and leaving the auto-suspension state. It is precisely the long-term existence of this mechanism that allows the TC to transport over long distances.

An increase in particle concentration can significantly increase the positive bed-parallel total force and positive bed-normal total force. The former improves the conversion rate of energy to particle kinetic energy in the TC system, whereby the current achieves an increase in advance velocity. The latter is generated by the increase of positive lift force, added mass force and contact force, which are essential to help to achieve and maintain particle auto-suspension. As a result, the increase in particle concentration drives the current to meet the auto-suspension criterion more easily. 
For different small slope conditions, the change in slope has an insignificant effect on both the depth-averaged auto-suspension index and the bed-parallel total force and bed-normal total force. Both negative effective drag force and positive lift force can be increased to some extent, but such enhancement is a mutually constrained balance. Additionally, the conversion rates of particle gravitational potential energy and fluid potential energy grow appreciably as the slope increases.

Particle entrainment (re-suspension) on the erodible terrain, provided with more prominent non-Newtonian dynamics, can provide more potential energy for TCs. The effect of this on particle auto-suspension investigated by fluid-particle coupled models, including particle transport intensity, system energy input and transformation, etc., is a work that is worth looking forward to and meaningful in the future.

\begin{acknowledgments}

\end{acknowledgments}

\backsection[Funding]
{The work was supported by the National Natural Science Foundation of China (Grant No. 12172331 for P. H., No. 12002334 for C. Z., No. 12072319 for Z. Y., and No. 12272344 for T. P.) and Zhejiang Provincial Natural Science Foundation (Grant No. LR19E090002 for P. H. and No. LQ21A020004 for C. Z.).}

\backsection[Declaration of interests]{The authors report no conflict of interest.}

\bibliographystyle{jfm}
%\bibliography{jfm}

%Use of the above commands will create a bibliography using the .bib file. Shown below is a bibliography built from individual items.

%% End of file `jfm2esam.bib'.

\end{document}